\documentclass[preprint,authoryear,12pt]{elsarticle}
\usepackage{graphicx}
\usepackage{amssymb,amsfonts,amsmath}
\usepackage{multirow}

\begin{document}

\newcommand{\EQ}{Eq.~}
\newcommand{\EQS}{Eqs.~}
\newcommand{\FIG}{Fig.~}
\newcommand{\FIGS}{Figs.~}
\newcommand{\TAB}{Table~}
\newcommand{\TABS}{Tables~}
\newcommand{\SEC}{Sec.~}
\newcommand{\SECS}{Secs.~}

\newcommand{\MbG}{\overline{b}_{\rm G}}
\newcommand{\MbB}{\overline{b}_{\rm B}}
\newcommand{\Ca}{\theta(1-\mu)+(1-\theta)\mu}
\newcommand{\Cb}{\theta\mu+(1-\theta)(1-\mu)}

\begin{frontmatter} 

\title{Ingroup favoritism and intergroup cooperation under indirect reciprocity based on group reputation}

\author[utokyo,jst]{Naoki Masuda\corref{cor1}
}
\address[utokyo]{Department of Mathematical Informatics,
The University of Tokyo,
7-3-1 Hongo, Bunkyo, Tokyo 113-8656, Japan}
\address[jst]{PRESTO, Japan Science and Technology Agency,
4-1-8 Honcho, Kawaguchi, Saitama 332-0012, Japan}
\cortext[cor1]{masuda@mist.i.u-tokyo.ac.jp}

\end{frontmatter} 

\newpage

\section*{Abstract}

Indirect reciprocity in which players cooperate with unacquainted other players having good reputations is a mechanism for cooperation in relatively large populations subjected to social dilemma situations. When the population has group structure, as is often found in social networks, players in experiments are considered to show behavior that deviates from existing theoretical models of indirect reciprocity. First, players often show ingroup favoritism (i.e., cooperation only within the group) rather than full cooperation (i.e., cooperation within and across groups), even though the latter is Pareto efficient. Second, in general, humans approximate outgroup members' personal characteristics, presumably including the reputation used for indirect reciprocity, by a single value attached to the group. Humans use such a stereotypic approximation, a phenomenon known as outgroup homogeneity in social psychology. I propose a model of indirect reciprocity in populations with group structure to examine the possibility of ingroup favoritism and full cooperation. In accordance with outgroup homogeneity, I assume that players approximate outgroup members' personal reputations by a single reputation value attached to the group. I show that ingroup favoritism and full cooperation are stable under different social norms (i.e., rules for assigning reputations) such that they do not coexist in a single model. If players are forced to consistently use the same social norm for assessing different types of interactions (i.e., ingroup versus outgroup interactions), only full cooperation survives. The discovered mechanism is distinct from any form of group selection. The results also suggest potential methods for reducing ingroup bias to shift the equilibrium from ingroup favoritism to full cooperation.

\newpage

\section{Introduction}\label{sec:introduction}

Humans and other animals often show cooperation in social dilemma
situations, in which defection apparently seems more lucrative than
cooperation. A main mechanism
governing cooperation in such situations is
direct reciprocity, in which the same pairs of players repeatedly
interact to realize mutual cooperation
\citep{Trivers1971,Axelrod1984book,Nowak2006book}. In fact,
individuals who do not repeatedly
interact also cooperate with others.
In this situation, reputation-based indirect reciprocity,
also known as downstream reciprocity, is a viable mechanism
for cooperation \citep{Nowak1998jtb,Leimar2001RoyalB,OhtsukiIwasa2004jtb,OhtsukiIwasa2007jtb,Nowak2005Nature,Brandt2005PNAS,Brandt2006jtb}.  In this mechanism, which I
refer to as indirect reciprocity for simplicity, individuals carry
their own reputation scores, which represent an evaluation of their
past actions toward others. Individuals are motivated to
cooperate to gain good reputations so that they are helped by
others in the future
or to reward (punish) good (bad) others.
Indirect reciprocity facilitates cooperation in a larger
population than in the case of direct reciprocity
because unacquainted players can cooperate with each other.
Although evidence of indirect reciprocity is relatively scarce
for nonhumans (but see \cite{Bshary2006Nature}),
it is widely accepted as explanation for
cooperation in humans \citep{Nowak2005Nature}.

Humans, in particular, belong to groups identified by traits,
such as age, ethnicity, and culture.  Individuals presumably interact
more frequently with ingroup than outgroup members. Group structure
has been a main topic of research in social
psychology and sociology for many decades
\citep{Brown2000book,Dovidio2005book} and in network science
\citep{Fortunato2010PhysRep}. Experimental evidence suggests that,
when the population of players has group structure,
two phenomena that are not captured by existing models of indirect reciprocity take place.

First,
in group-structured populations,
humans
\citep{Sedikides1998book,Brewer1999JSI,Hewstone2002ARP,Dovidio2005book,Efferson2008Science} 
and even insect larvae \citep{Lize2006RoyalB}
show various forms of ingroup favoritism.
In social dilemma games, individuals behave more cooperatively toward ingroup
than outgroup members (e.g.,
\cite{DeCremer1999EJSP,Goette2006AER,Fowler2007JPoli,Rand2009PNAS,Yamagishi1998AJSP,Yamagishi1999chapter,YamagishiMifune2008RS}). 
Ingroup favoritism in social dilemma situations may occur as a result of
indirect reciprocity confined in the group \citep{Yamagishi1998AJSP,Yamagishi1999chapter,YamagishiMifune2008RS}. In contrast,
ingroup favoritism in social
dilemma games is not Pareto efficient because 
individuals would receive larger payoffs if they also cooperated
across groups. 
Under what conditions are ingroup favoritism and intergroup cooperation
sustained by indirect reciprocity? Can they bistable?

Ingroup favoritism, which has also been analyzed in the context of
tag-based cooperation, the green beard effect, and the armpit effect,
has been considered to be a theoretical challenge
(e.g., \cite{Antal2009PNAS}). Nevertheless, recent research has
revealed their
mechanisms, including the loose coupling of altruistic trait and tag in inheritance \citep{Jansen2006Nature}, a relatively fast mutation that simultaneously changes strategy and tag \citep{Traulsen2007PlosOne,Traulsen2008EPJB}, a tag's relatively fast mutation as compared to the strategy's mutation \citep{Antal2009PNAS}, conflicts between groups \citep{ChoiBowles2007Science,Garcia2011EHB}, partial knowledge of others' strategies \citep{MasudaOhtsuki2007RoyalB}, and gene-culture coevolution \citep{Ihara2011PhilB}. However, indirect reciprocity accounts for ingroup favoritism, as is relevant to previous experiments
\citep{Yamagishi1998AJSP,Yamagishi1999chapter,YamagishiMifune2008RS} is lacking.

Second, in a population with group structure, individuals tend to
approximate outgroup individuals' characteristics by a single value
attached to the group.  This type of stereotype is known as outgroup
homogeneity in social psychology
\citep{Jones1981PSPB,Ostrom1992PsychBull,Sedikides1998book,Brown2000book},
and it posits that outgroup
members tend to be regarded to resemble each other more
than they actually do.
It is also reasonable from the viewpoint of cognitive burden of remembering
each individual's properties that humans generally 
resort to outgroup homogeneity.
Therefore, in indirect reciprocity games in group structured populations,
it seems to be natural to assume outgroup homogeneity. In other words,
individuals may not care about or have access to personal reputations
of those in different groups and 
approximate an outgroup individual's reputation by a group reputation.

Some previous models analyzed the situations in which
players do not have access to individuals' reputations.
This is simply because it may be
difficult for an individual in a large population
to separately keep track of other people's reputations
even if gossiping helps dissemination of information.
This case of incomplete information has been theoretically modeled by
introducing the probability that an individual sees others' reputations
in each interaction
\citep{Nowak1998nature,Nowak1998jtb,Brandt2005PNAS,Brandt2006jtb,SuzukiToquenaga2005JTB,NakamuraMasuda2011PLoSComputBiol}.
However, these studies do not have to do with the approximation of 
individuals' personal reputations by group reputations.

By analyzing a model of an indirect reciprocity game based on group
reputation, I provide an indirect reciprocity account for ingroup
favoritism for the first time. In addition, through an exhaustive search, I identify
all the different types of stable homogeneous populations that yield
full cooperation (intragroup and
intergroup cooperation) or ingroup favoritism.

\section{Methods}

\subsection{Model}

\subsubsection{Population structure and the donation game}

I assume that the population is composed of infinitely many groups
each of which is of infinite size. Each player belongs to one group.

Players are involved in a series of the donation game, which is
essentially a type
of prisoner's dilemma game. In each round,
a donor and recipient are selected from the population in a completely
random manner.  Each player is equally likely to be selected as donor or recipient.
The donor may refer to the recipient's reputation
and select one of the two actions, cooperation (C) or defection
(D). If the donor cooperates, the donor pays
cost $c>0$, and the recipient receives benefit $b (>c)$. If
the donor defects, the payoffs to the donor and recipient are
equal to 0. 
Because the roles are asymmetric in a single game,
the present game differs from the one-shot or standard
iterated versions of the prisoner's dilemma game. This game is widely
used for studying mechanisms for cooperation including indirect reciprocity
\citep{Nowak2005Nature,Nowak2006book,Nowak2006Science}.

Rounds are repeated a sufficient number of times with different pairs of
donors and recipients. Because the population is infinite, 
no pair of players meets more than once, thereby avoiding
the possibility of direct
reciprocity (e.g., \cite{Nowak1998jtb,OhtsukiIwasa2004jtb}). The payoff to each
player is defined as the average payoff per round.

The groups to which the donor and recipient belong
are denoted by $g_{\rm d}$ and $g_{\rm r}$, 
respectively.
The simultaneously selected donor and recipient
belong to the same group with probability $r^{\rm in}$ (i.e., $g_{\rm d}=g_{\rm r}$;
\FIG\ref{fig:ingroup outgroup observers}A) and
different groups with probability $r^{\rm out}\equiv 1-r^{\rm in}$
(i.e., $g_{\rm d}\neq g_{\rm r}$; \FIG\ref{fig:ingroup outgroup observers}B).

\subsubsection{Social norms}

At the end of each round, observers assign binary reputations, good (G) or bad (B),
to the donor and donor's group ($g_{\rm d}$) according to a given
social norm.  I consider up
to so-called second-order social norms with which the observers
assign G or B as a function of the donor's action and the reputation
(i.e., G or B) of the recipient or recipient's group ($g_{\rm r}$).
Representative second-order social norms are shown in
\FIG\ref{fig:norms}. Under image scoring (``scoring'' in \FIG\ref{fig:norms}), an observer
regards a donor's action C or D to be G or B, respectively, regardless of
the recipient's reputation. In the absence of a group-structured population, scoring does not
realize cooperation based on indirect reciprocity unless certain specific
conditions are met \citep{Nowak1998jtb,Brandt2005PNAS,Brandt2006jtb,Leimar2001RoyalB,OhtsukiIwasa2004jtb}. Simple standing (``standing'' in \FIG\ref{fig:norms}),
and stern judging (``judging'' in \FIG\ref{fig:norms}; also known as Kandori)
enable full cooperation
\citep{Leimar2001RoyalB,OhtsukiIwasa2004jtb}. Shunning also enables full cooperation
if the players' reputations are initially C and the number of
rounds is finite \citep{OhtsukiIwasa2007jtb} or
if the players' reputations are partially invisible \citep{NakamuraMasuda2011PLoSComputBiol}.

\begin{figure}[h]
\begin{center}
\includegraphics[width=10cm]{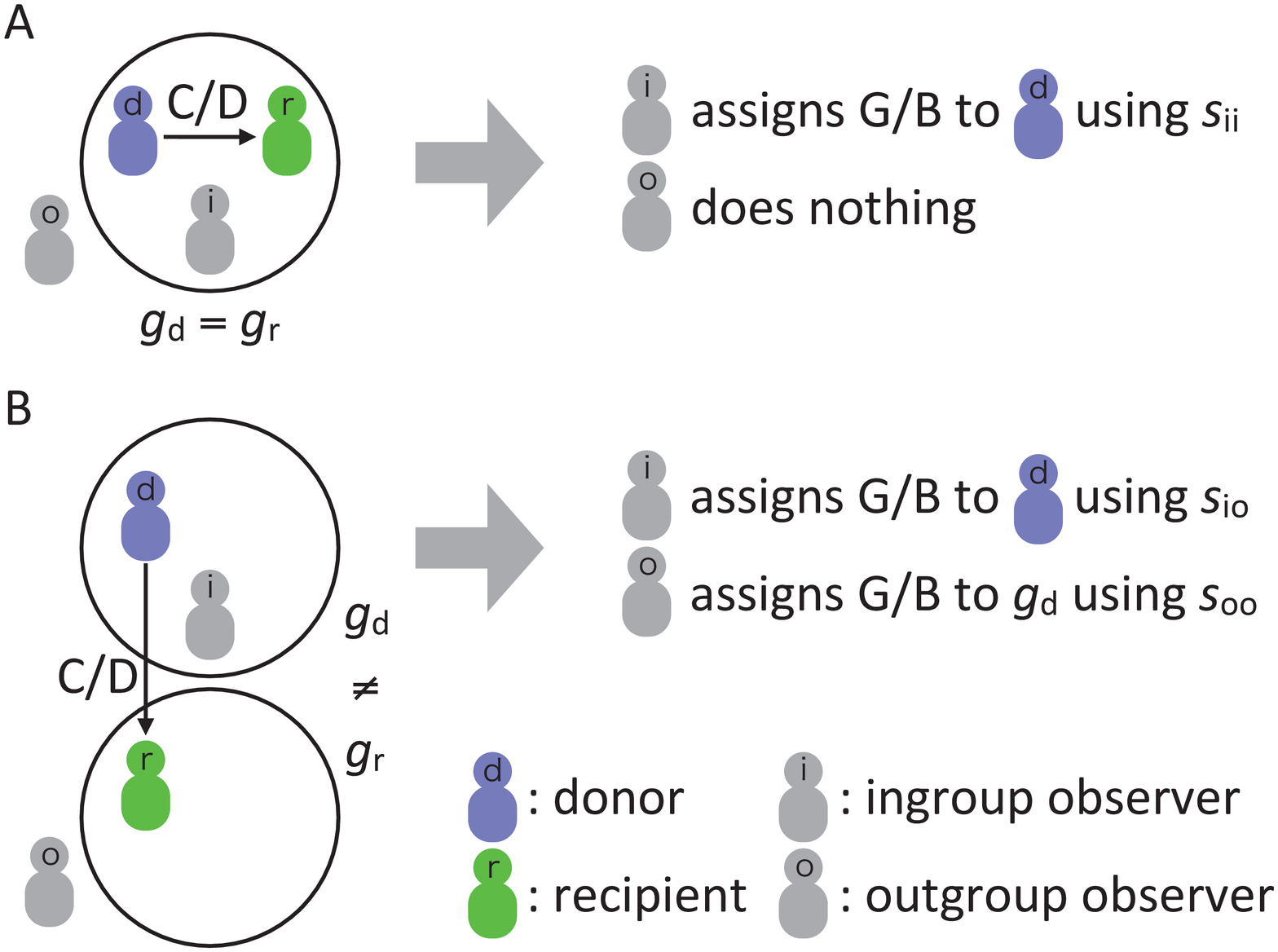}
\caption{Schematic representation of ingroup and outgroup
  observers. In A, the donor's group $g_{\rm d}$ and the
  recipient's group $g_{\rm r}$ are identical. This event occurs with
  probability $r^{\rm in}$. In B, $g_{\rm d}\neq g_{\rm
    r}$. This event occurs with probability $r^{\rm out}=1-r^{\rm
    in}$.}
\label{fig:ingroup outgroup observers}
\end{center}
\end{figure}

\begin{figure}[h]
\begin{center}
\includegraphics[width=7cm]{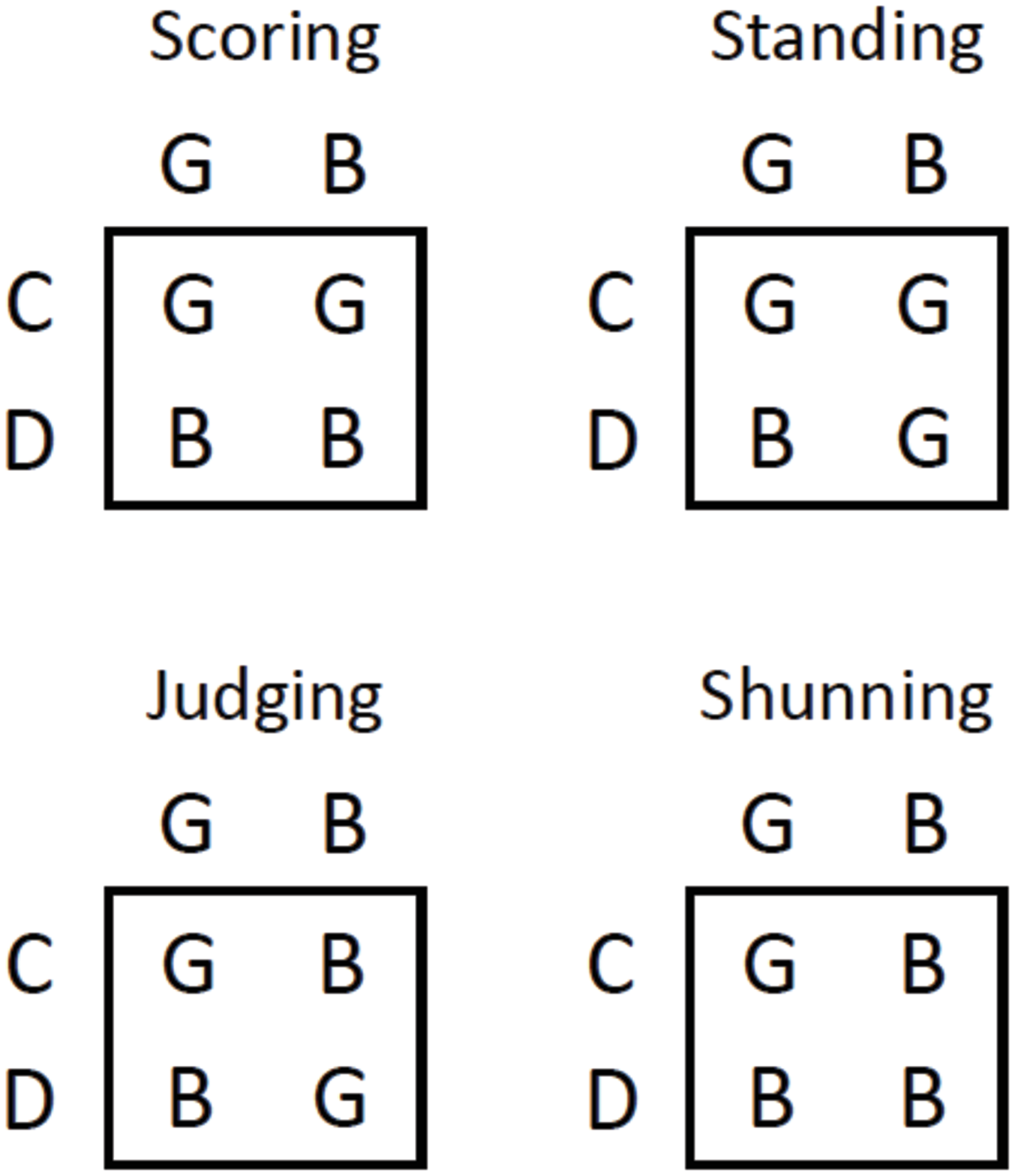}
\caption{Typical second-order social norms. The rows outside the boxes
  represent the donor's actions (C or D), and the columns
  represent the recipient's reputations (G or B). The entries inside the boxes represent the reputations that the observer assigns to the donor in each case.}
\label{fig:norms}
\end{center}
\end{figure}

In the presence of group structure,
four possible locations of the observer are
schematically shown in
\FIG\ref{fig:ingroup outgroup observers}.
I call the observer belonging to $g_{\rm d}$ an ``ingroup'' observer.
Otherwise, the observer is called an ``outgroup'' observer.

The observers can adopt different social norms
for the four cases, as summarized in \FIG\ref{fig:ingroup outgroup observers}.
When the donor and recipient belong to the same group
(\FIG\ref{fig:ingroup outgroup observers}A), the ingroup observer uses
the norm
denoted by $s_{\rm ii}$ to update the donor's personal reputation.
In this situation,
the outgroup observer does not update the donor's or $g_{\rm d}$'s
reputation (but see Appendix~\ref{sec:appendix variant}).
When the donor and recipient belong to different groups
(\FIG\ref{fig:ingroup outgroup observers}B), the ingroup
observer uses the norm denoted by $s_{\rm io}$ to update the donor's
personal reputation.
In this situation,
the outgroup observer uses the norm denoted by
$s_{\rm oo}$ to update $g_{\rm d}$'s reputation. 
These four cases are explained in more detail in \SEC\ref{sub:update}.

The distinction between $s_{\rm ii}$ and $s_{\rm io}$ allows the ingroup observer to use a double standard for assessing donors. For example, a donor defecting against an ingroup G recipient may be regarded to be B, whereas a defection against an outgroup G recipient may be regarded as G. Such different assessments would not be allowed if $s_{\rm ii}$ and $s_{\rm io}$ are not distinguished.

I call $s_{\rm ii}$, $s_{\rm io}$, and $s_{\rm oo}$ subnorms.
All the players are assumed to share the subnorms.
The typical norms shown in \FIG\ref{fig:norms} can be used as subnorms.
A subnorm is specified by assigning G or B to each combination of
the donor's action (i.e., C or D) and recipient's reputation (i.e., G or
B). Therefore, there are $2^4=16$ subnorms. 
An entire social norm of a population consists of
a combination of the three subnorms, and there are $16^3=4096$ social norms.

\subsubsection{Action rule}

The action rule refers to the mapping from the recipient's reputation
(i.e., G or B) to the donor's action (i.e., C or D).
The AllC and AllD donors cooperate and defect, respectively, regardless of the recipient's
reputation. A discriminator (Disc) 
donor cooperates or defects when the recipient's
reputation is G or B, respectively.
An anti-discriminator (AntiDisc) donor cooperates or defects
when the recipient's reputation is B or G, respectively.

The donor is allowed to use different action rules toward ingroup and outgroup
recipients. For example, a donor who adopts AllC and AllD 
toward ingroup and outgroup recipients, respectively, implements
reputation-independent ingroup favoritism.
There are $4\times 4=16$ action rules.
A donor
refers to the recipient's personal reputation when $g_{\rm d}=g_{\rm
r}$ (\FIG\ref{fig:ingroup outgroup
observers}A) and to $g_{\rm r}$'s group reputation when $g_{\rm d}\neq g_{\rm
r}$ (\FIG\ref{fig:ingroup outgroup observers}B). 

\subsubsection{Reputation updates}\label{sub:update}

In each round, the ingroup and outgroup observers update the donor's
and $g_{\rm d}$'s reputations, respectively.

If $g_{\rm d}=g_{\rm r}$,
the donor is assumed to recognize the recipient's personal reputation
(\FIG\ref{fig:ingroup outgroup observers}A).
An ingroup observer in this situation
updates the donor's personal reputation on the basis of the donor's action,
the recipient's personal reputation, and subnorm $s_{\rm ii}$.
An outgroup observer in this situation
is assumed not to update $g_{\rm d}$'s reputation
because such an observer does not know the recipient's personal
reputation, although the donor does. Then, the outgroup observer may
want to refrain from evaluating the donor because the donor and the observer use different information about the recipient.
I also analyzed a variant of the model in which the outgroup observer updates 
$g_{\rm d}$'s reputation in this situation. The results are roughly the same as those obtained for the original model (Appendix~\ref{sec:appendix variant}).

If $g_{\rm d}\neq g_{\rm r}$, the donor is assumed to
recognize $g_{\rm r}$'s reputation, but not the recipient's personal reputation
(\FIG\ref{fig:ingroup outgroup observers}B).
An ingroup observer in this situation
updates the donor's personal reputation on the basis of the donor's action,
$g_{\rm r}$'s reputation, and subnorm $s_{\rm io}$.
Both the donor and observer
refer to $g_{\rm r}$'s reputation and not to the recipient's personal reputation.
An outgroup observer in this situation
updates $g_{\rm d}$'s reputation based on the 
donor's action, $g_{\rm r}$'s reputation, and subnorm $s_{\rm oo}$.

An outgroup observer 
knows the recipient's personal reputation if the
observer and recipient are in the same group. However, the observer is
assumed to ignore this information 
for two reasons. First,
it is evident for the observer that the
donor does not have access to the recipient's personal reputation.
To explain the second reason, let us
consider an outgroup observer who belongs to
$g_{\rm r}$ in a certain round. Assume that
this observer assigns
a new reputation to $g_{\rm d}$ according to
a subnorm different from one used when the observer does not belong to
$g_{\rm r}$. The same observer does not belong to $g_{\rm r}$ when the observer
updates the $g_{\rm d}$'s group reputation next time.
This is because the probability that the observer belongs to $g_{\rm r}$ is
infinitesimally small because of the assumption of infinite groups.
Therefore, the subnorm used
when the observer belongs to $g_{\rm r}$ is rarely used and 
immaterial in the present model.

Finally, observers commit
reputation assessment error. With probability $\epsilon$, ingroup
and outgroup observers independently assign
the reputation opposite to the intended one 
to the donor and $g_{\rm d}$, respectively.
I introduce this error because G and B players must coexist in the population to distinguish the payoff values for different pairs of action rule and social norm (action--norm pair); such a distinction is necessary for the stability analysis in the following discussion. For simplicity, I neglect other types of error. 

\subsubsection{Mutant types}\label{sub:type of mutants}

To examine the stability of an action rule under a given social norm,
I consider two types of mutants. 

The first is a single mutant
that invades a group. There are $16-1=15$ types of single mutants.
A single mutant does not
affect the action rule, norm, or reputation of the group that the mutant
belongs to because of the assumption of infinite group size.

The second type is a group mutant.  
A homogeneous group composed of mutants
may make the mutant type stronger than the resident type. For example, a group composed of players 
who cooperate with ingroup recipients and defect against outgroup recipients may invade
a fully cooperative population if any
intergroup interaction (i.e., C or D) is regarded to be G under $s_{\rm oo}$.
By definition, a
group mutant is a homogeneous group of mutants that is different from
the resident players in either the action rule or social norm. I
consider two varieties of group mutants, as described in
\SEC\ref{sec:results}. 

\subsection{Analysis methods}\label{sec:analysis}

\subsubsection{Reputation scores in the equilibrium}

Consider a homogeneous resident population in which all players
share an action--norm pair. I will examine the stability of this population against
invasion by single and group mutants. For this purpose, I calculate
the fraction of players with a G reputation, probability of cooperation, and payoff after
infinitely many rounds.

Denote by $p^*$ and $p_{\rm g}^*$
the equilibrium probabilities that the player's and group's reputations are
G, respectively.
The self-consistent equation for $p^*$ is given by
\begin{equation}
p^* = r^{\rm in}\left[ p^* \Phi_{\rm G}^{\rm in}(\sigma^{\rm in}) + (1-p^*)
\Phi_{\rm B}^{\rm in}(\sigma^{\rm in}) \right]
+ r^{\rm out}\left[ p_{\rm g}^* \Phi_{\rm G}^{\rm in}(\sigma^{\rm out}) +
  (1-p_{\rm g}^*)
\Phi_{\rm B}^{\rm in}(\sigma^{\rm out}) \right],
\label{eq:p equilibrium}
\end{equation}
where $\sigma^{\rm in}$ and $\sigma^{\rm out}$ are the action rules
(i.e., AllC, Disc, AntiDisc, or AllD) that the donor adopts toward
ingroup and outgroup recipients, respectively.  $\Phi_{\rm G}^{\rm
  in}(\sigma^{\rm in})$ and $\Phi_{\rm B}^{\rm in}(\sigma^{\rm in})$
are the probabilities that the ingroup observer, based on $s_{\rm
  ii}$, assigns reputation G to a donor who has played with a G or B
ingroup recipient (i.e., $g_{\rm d}=g_{\rm r}$), respectively
(\FIG\ref{fig:ingroup outgroup observers}A).
Similarly $\Phi_{\rm G}^{\rm in}(\sigma^{\rm out})$
and $\Phi_{\rm B}^{\rm in}(\sigma^{\rm out})$ apply when the recipient
is in a different group (i.e., $g_{\rm d}\neq g_{\rm r}$) and the
observer uses $s_{\rm io}$ (\FIG\ref{fig:ingroup outgroup
  observers}B).  It should be noted that $\Phi_{\rm G}^{\rm
  in}(\sigma^{\rm in})$ and $\Phi_{\rm G}^{\rm in}(\sigma^{\rm out})$,
for example, may differ from each other even if $\sigma^{\rm in} = \sigma^{\rm
  out}$. Owing to the reputation assignment error, $\Phi_{\rm G}^{\rm
  in}(\sigma^{\rm in})$, $\Phi_{\rm B}^{\rm in}(\sigma^{\rm in})$,
$\Phi_{\rm G}^{\rm in}(\sigma^{\rm out})$, $\Phi_{\rm B}^{\rm
  in}(\sigma^{\rm out})$ $\in \{\epsilon,1-\epsilon\}$ holds true.
For example, if the donor is Disc toward ingroup recipients and
subnorm $s_{\rm ii}$ is scoring, $\Phi_{\rm G}^{\rm
  in}(\sigma^{\rm in})=1-\epsilon$ and $\Phi_{\rm B}^{\rm
  in}(\sigma^{\rm in})=\epsilon$.

The self-consistent equation for $p_{\rm g}^*$ is given by
\begin{equation}
p_{\rm g}^* = r^{\rm in}p_{\rm g}^* + r^{\rm out} \left[
p_{\rm g}^* \Phi_{\rm G}^{\rm out}(\sigma^{\rm out})
+ (1-p_{\rm g}^*) \Phi_{\rm B}^{\rm out}(\sigma^{\rm out})
 \right],
\label{eq:pg equilibrium case A}
\end{equation}
where $\Phi_{\rm G}^{\rm out}(\sigma^{\rm out})\in
\{\epsilon,1-\epsilon\}$ and $\Phi_{\rm B}^{\rm out}(\sigma^{\rm
  out})\in \{\epsilon,1-\epsilon\}$ are the probabilities that the
outgroup observer, based on $s_{\rm oo}$, assigns reputation G to
the donor's group when the donor has played with a G or B
outgroup recipient (i.e., $g_{\rm d}\neq g_{\rm r}$), respectively
(\FIG\ref{fig:ingroup outgroup observers}B).
The first term on the right-hand side of \EQ\eqref{eq:pg equilibrium case A}
corresponds to the fact that $g_{\rm d}$'s reputation is not updated in the
situation illustrated in \FIG\ref{fig:ingroup
  outgroup observers}A.

Equations~\eqref{eq:p equilibrium} and \eqref{eq:pg equilibrium case
  A} lead to
\begin{equation}
p^* = \frac{r^{\rm in}\Phi_{\rm B}^{\rm in}(\sigma^{\rm in}) + r^{\rm
    out}\left[ p_{\rm g}^* \Phi_{\rm G}^{\rm in}(\sigma^{\rm out})
+ (1-p_{\rm g}^*)\Phi_{\rm B}^{\rm in}(\sigma^{\rm out}) \right]}
{1-r^{\rm in}\Phi_{\rm G}^{\rm in}(\sigma^{\rm in})+r^{\rm in}\Phi_{\rm B}^{\rm
  in}(\sigma^{\rm in})}
\end{equation}
and
\begin{equation}
p_{\rm g}^* = \frac{r^{\rm out}\Phi_{\rm B}^{\rm out}(\sigma^{\rm out})}
{1-r^{\rm in}-r^{\rm out}\Phi_{\rm G}^{\rm out}(\sigma^{\rm out})+r^{\rm
    out}\Phi_{\rm B}^{\rm out}(\sigma^{\rm out})}.
\end{equation}

\subsubsection{Stability against invasion by single mutants}

To examine the stability of the action rule ($\sigma^{\rm in}$,
$\sigma^{\rm out}$) against invasion by single mutants under a given social
norm, I consider a single mutant with action rule ($\sigma^{\rm in
  \prime}$, $\sigma^{\rm out \prime}$). 
Because the group is
assumed to be infinitely large, a single mutant does not
change the reputation of the invaded group. The equilibrium
probability $p^{\prime *}$ that a mutant receives personal reputation G
is given by
\begin{equation}
p^{\prime *} = r^{\rm in}\left[ p^*\Phi_{\rm G}^{\rm in}(\sigma^{\rm in \prime})
+(1-p^*)\Phi_{\rm B}^{\rm in}(\sigma^{\rm in \prime}) \right]
+ r^{\rm out}\left[ p_{\rm g}^*\Phi_{\rm G}^{\rm in}(\sigma^{\rm out \prime})
+ (1-p_{\rm g}^*)\Phi_{\rm B}^{\rm in}(\sigma^{\rm out \prime})\right].
\label{eq:p'*}
\end{equation}

When the probability that the donor and $g_{\rm d}$ have a G reputation is equal to $p$ and
$p_{\rm g}$, respectively,
the resident donor cooperates with probability
\begin{equation}
r^{\rm in}\Psi(\sigma^{\rm in}, p) + r^{\rm out}\Psi(\sigma^{\rm out}, p_{\rm g}),
\label{eq:prob C}
\end{equation}
where
\begin{equation}
\Psi(\tilde{\sigma}, \tilde{p})=
\tilde{p}\zeta_{\rm G}(\tilde{\sigma})+(1-\tilde{p})\zeta_{\rm B}(\tilde{\sigma})\quad
(\tilde{p}=p, p_{\rm g})
\label{eq:Psi}
\end{equation}
is the probability that a donor with action rule $\tilde{\sigma}\in \{
\mbox{AllC, Disc, AntiDisc, AllD} \}$ 
cooperates when the recipient's
personal or group reputation is G with probability $\tilde{p}$.
$\zeta_{\rm G}(\tilde{\sigma})$ and $\zeta_{\rm B}(\tilde{\sigma})$
($\tilde{\sigma}=\sigma^{\rm in}$ or $\sigma^{\rm out}$) are the probabilities
that a $\tilde{\sigma}$ donor cooperates with a G and B recipient, respectively.
AllC, Disc, AntiDisc, and AllD correspond to
$(\zeta_{\rm G}(\tilde{\sigma}), \zeta_{\rm B}(\tilde{\sigma})) = (1,1),
(1,0), (0,1)$, and $(0, 0)$, respectively.

The payoff to a resident ($\sigma^{\rm in}$, $\sigma^{\rm out}$)--player is given by
\begin{equation}
\pi = -c \left[r^{\rm in}\Psi(\sigma^{\rm in},p^*)+r^{\rm
    out}\Psi(\sigma^{\rm out},p_{\rm g}^*) \right] + b \left[r^{\rm in}\Psi
(\sigma^{\rm in}, p^*) + r^{\rm out}\Psi(\sigma^{\rm out}, p_{\rm g}^*) \right].
\label{eq:pi}
\end{equation}
The payoff to a ($\sigma^{\rm in\prime}$, $\sigma^{\rm out\prime}$)--mutant invading the homogeneous population of the resident action--norm pair is given by
\begin{equation}
\pi^{\prime} = -c \left[r^{\rm in}\Psi(\sigma^{\rm in
    \prime},p^*) + r^{\rm out}\Psi(\sigma^{\rm out\prime}, p_{\rm
    g}^*) \right] + b\left[ r^{\rm in}\Psi(\sigma^{\rm in}, p^{\prime
    *}) + r^{\rm out}\Psi(\sigma^{\rm out}, p_{\rm g}^*) \right].
\label{eq:pi'}
\end{equation}
If $\pi>\pi^{\prime}$ for any mutant, the pair of the action rule
($\sigma^{\rm in}$, $\sigma^{\rm out}$) and social norm
($s_{\rm ii}$, $s_{\rm io}$, $s_{\rm oo}$) is stable against invasion
by single mutants.

\subsubsection{Stability against invasion by group mutants}

For a mutant group composed of players sharing an action--norm pair, let
$p_{\rm g}^{\prime *}$ denote the equilibrium probability that the mutant group
has group reputation G.
I obtain
\begin{align}
p^{\prime *} =& r^{\rm in}\left[ p^{\prime *}\Phi_{\rm G}^{\rm
    in\prime}(\sigma^{\rm in\prime}) + (1-p^{\prime *})\Phi_{\rm B}^{\rm
    in\prime}(\sigma^{\rm in\prime})\right]\notag\\
&+ r^{\rm out}\left[ p_{\rm g}^*\Phi_{\rm G}^{\rm
    in\prime}(\sigma^{\rm out\prime}) + (1-p_{\rm g}^*)\Phi_{\rm B}^{\rm
    in\prime}(\sigma^{\rm out\prime})\right]
\label{eq:p'* group mutation}
\end{align}
and
\begin{equation}
p_{\rm g}^{\prime *} = r^{\rm in} p_{\rm g}^{\prime *} + r^{\rm out}\left[
p_{\rm g}^*\Phi_{\rm G}^{\rm out}(\sigma^{\rm out\prime}) + 
(1-p_{\rm g}^*)\Phi_{\rm B}^{\rm out}(\sigma^{\rm out\prime})
\right],\label{eq:pg'* group mutation case A}
\end{equation}
where $\Phi_{\rm G}^{\rm in\prime}(\sigma^{\rm in\prime})$ or
$\Phi_{\rm B}^{\rm in\prime}(\sigma^{\rm in\prime})$
is the
probability that an ingroup observer assigns reputation G to
a mutant donor who has played with a G or B
ingroup recipient, respectively. Even if
$\sigma^{\rm in\prime}$ and $\sigma^{\rm in}$ are the same,
$\Phi_{\rm G}^{\rm in\prime}(\sigma^{\rm in\prime})$ will be generally different from
$\Phi_{\rm G}^{\rm in}(\sigma^{\rm in})$ because the
ingroup observer in the mutant group may use a
subnorm $s_{\rm ii}$ that is different from one used in the
resident population. Parallel definitions apply to 
$\Phi_{\rm G}^{\rm in\prime}(\sigma^{\rm out\prime})$ and
$\Phi_{\rm B}^{\rm in\prime}(\sigma^{\rm out\prime})$.
Equations~\eqref{eq:p'* group mutation} and \eqref{eq:pg'* group
mutation case A}
yield
\begin{equation}
p^{\prime *} = \frac{r^{\rm in}\Phi_{\rm B}^{\rm
    in\prime}(\sigma^{\rm in\prime})+r^{\rm out}\left[p_{\rm g}^*\Phi_{\rm G}^{\rm
    in\prime}(\sigma^{\rm out\prime})+(1-p_{\rm g}^*)\Phi_{\rm B}^{\rm
    in\prime}(\sigma^{\rm out\prime})\right]}
{1-r^{\rm in}\Phi_{\rm G}^{\rm in\prime}(\sigma^{\rm in\prime})+r^{\rm
    in}\Phi_{\rm B}^{\rm in\prime}(\sigma^{\rm in\prime})}
\label{eq: p'* group mutation final}
\end{equation}
and
\begin{equation}
p_{\rm g}^{\prime *} = p_{\rm g}^* \Phi_{\rm G}^{\rm
  out}(\sigma^{\rm out\prime})
+ (1-p_{\rm g}^*)\Phi_{\rm B}^{\rm out}(\sigma^{\rm out\prime}),
\label{eq: pg'* group mutation final case A}
\end{equation}
respectively.

The payoff to a mutant player in the mutant group is given by
\begin{equation}
\pi_{\rm g}^{\prime} = -c \left[r^{\rm in}\Psi(\sigma^{\rm in\prime},
  p^{\prime *}) + r^{\rm out}\Psi(\sigma^{\rm out\prime},
p_{\rm g}^*)\right] + b \left[r^{\rm in}\Psi(\sigma^{\rm in\prime},
p^{\prime *})
+ r^{\rm out}\Psi(\sigma^{\rm out}, p_{\rm g}^{\prime *})\right].
\label{eq:pi g'}
\end{equation}
If $\pi>\pi_{\rm g}^{\prime}$ holds true for any group mutant player,
the resident population is stable
against invasion by group mutants.

\section{Results}\label{sec:results}

\subsection{Action--norm pairs stable against invasion by single
  mutants}\label{sub:against single mutants}

\begin{figure}[h]
\begin{center}
\includegraphics[width=10cm]{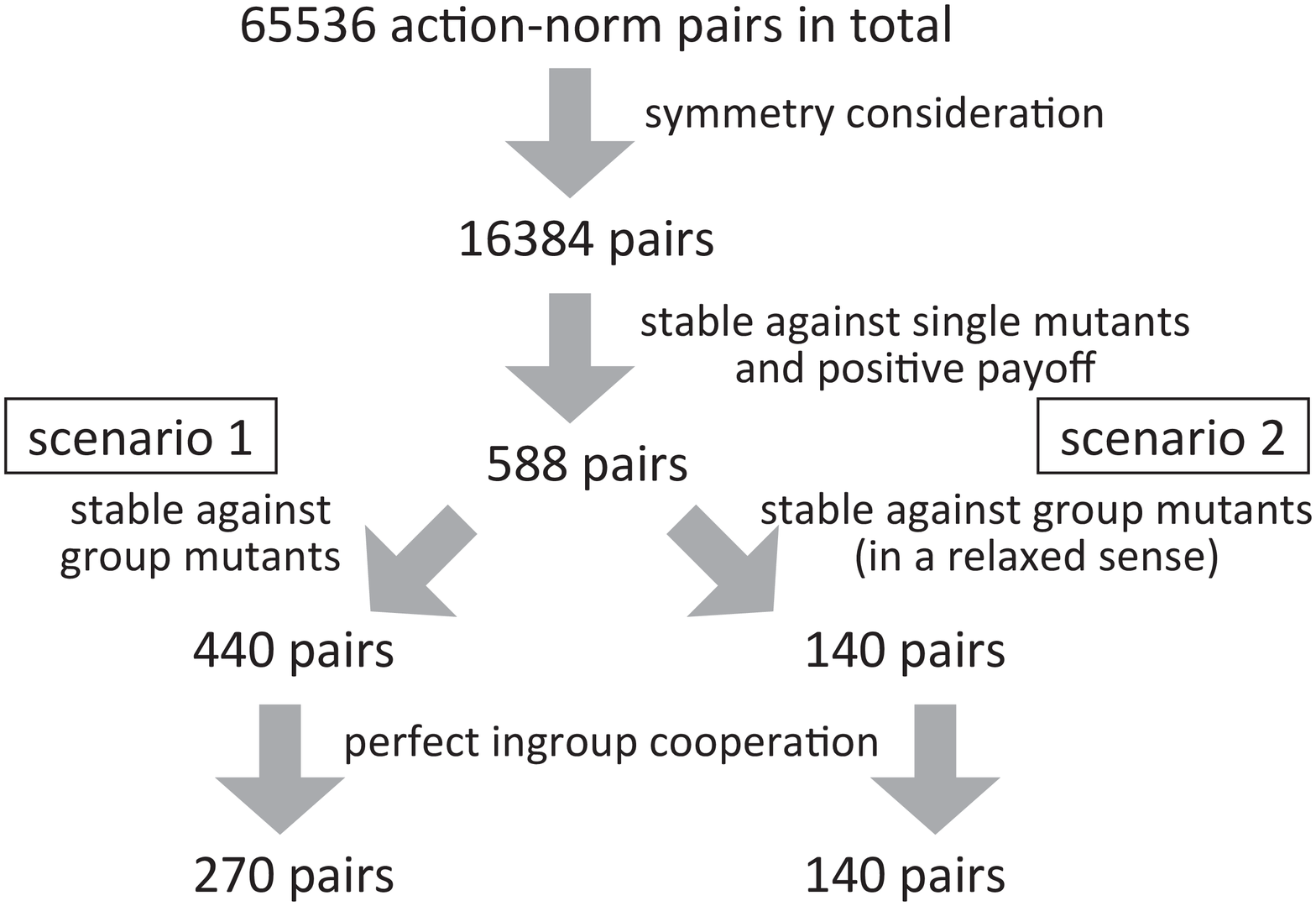}
\caption{Procedure for obtaining
the stable action--norm pairs with perfect ingroup cooperation shown in 
\TABS\ref{tab:stable pairs 1} and \ref{tab:stable pairs 2}.}
\label{fig:equilibrium selection}
\end{center}
\end{figure}

There are 16 action rules and $16^3=4096$ social norms,
which leads to $16\times 4096 = 65536$ action--norm pairs.
Because of the symmetry with respect to the swapping of G and B,
I neglect action--norm pairs in which the action rule
(i.e., AllC, Disc, AntiDisc, or AllD) toward ingroup recipients
is $\sigma^{\rm
  in}=$ AntiDisc without loss of generality. 
Such an action--norm pair can be converted to $\sigma^{\rm in}=$ Disc
by swapping G and B in the action rule and social norm.
The model is also invariant if G and B group reputations
are completely swapped in the action rule toward outgroup recipients
$\sigma^{\rm out}$ and subnorms $s_{\rm io}$ and $s_{\rm oo}$.
Therefore, I can also neglect the action--norm pairs with
$\sigma^{\rm out}=$ AntiDisc without loss of generality.
This symmetry consideration leaves $65536/4=16384$ action--norm pairs
(\FIG\ref{fig:equilibrium selection}).

I exhaustively examined the stability of all $16\times 4096 = 65536$
action--norm pairs. A similar exhaustive search was first conducted in
\citep{OhtsukiIwasa2004jtb} for an indirect reciprocity game without
group structure in the population. In the following, $\pi$
  (\EQ\eqref{eq:pi}) mentions the player's payoff in the resident
population in the limit of no reputation assignment error, i.e.,
$\epsilon\to 0$.

I first describe action rules that are stable against invasion by single mutants
under a given social norm.
I identified them using \EQS\eqref{eq:p equilibrium}--\eqref{eq:pi'}.
Under any given social norm, action rule ($\sigma^{\rm in}$, $\sigma^{\rm out}$) $=$
(AllD, AllD) is stable and yields $\pi=0$.
Other action--norm pairs also yield $\pi=0$, but
there are
588 stable action--norm pairs with $\pi>0$
(\FIG\ref{fig:equilibrium selection}). For a given social norm,
at most one action rule that yields a positive payoff is stable.
For all 588 solutions, the condition for stability against
invasion by single mutants (i.e., $\pi>\pi^{\prime}$, where $\pi$ and $\pi^{\prime}$ are given by \EQS\eqref{eq:pi} and \eqref{eq:pi'}, respectively) is given by
\begin{equation}
b r^{\rm in}>c.
\label{eq:b rin > c}
\end{equation}
Equation~\eqref{eq:b rin > c} implies that
cooperation is likely when the benefit-to-cost ratio is large, which
is a standard result for different mechanisms of cooperation in social
dilemma games \citep{Nowak2006Science}.
Cooperation is also likely when intragroup interaction is relatively
more frequent than intergroup interaction (i.e., large $r^{\rm in}$).

\subsection{Stability against invasion by group mutants}

The stability of these 588 action--norm pairs
against invasion by group mutants was also examined based on \EQS\eqref{eq:p'* group mutation}--\eqref{eq:pi g'}. Properly setting the variety 
of group mutants is not a
trivial issue. At most,
$65536 -1 = 65535$ types of group mutants that differ from the
resident population in either action rule or social norm are possible.
However, an arbitrarily selected homogeneous mutant group may be fragile to 
invasion by different single mutants into the mutant group.
Although I do not model evolutionary dynamics,
evolution would not allow
the emergence and maintenance of such weak mutant groups. With this in mind, I consider
two group mutation scenarios.

\subsubsection{Scenario 1}\label{sub:scenario 1}

Single mutants may invade the resident population when
\EQ\eqref{eq:b rin > c} is violated.
In this scenario 1,
the mutants are assumed to differ 
from the resident population in the action rule, but not the social
norm, for simplicity.
There are $16-1=15$ such mutants, and
some of them, including ($\sigma^{\rm in}$, $\sigma^{\rm out}$) $=$
(AllD, AllD), can invade the resident population when $1<b/c<1/r^{\rm in}$.
Such mutant action rules may spread to occupy a single group
when \EQ\eqref{eq:b rin > c} is violated.
I consider the stability of the resident
population against the homogeneous groups of mutants that invade the resident population 
as single mutants when $1<b/c<1/r^{\rm in}$.

Among the 588 action--norm pairs that yield $\pi>0$,
440 pairs are
stable against group mutation.
Among these 440 pairs, I focus on those yielding perfect intragroup
cooperation, i.e., those yielding $\lim_{\epsilon\to
  0}\Psi(\sigma^{\rm in},p^*)=1$, where $\Psi$ and $p^*$ are given in \SEC\ref{sec:analysis}. For the other stable pairs, see Appendix~\ref{sec:outgroup favoritism}.
This criterion is satisfied by 270 pairs (\FIG\ref{fig:equilibrium selection}).
For all 270 pairs,
every player obtains personal reputation G
(i.e., $\lim_{\epsilon\to 0}p^*=1$), and the donor cooperates with ingroup
recipients because the recipients have reputation G
(i.e., $\sigma^{\rm in}=$ Disc).

In all 270 pairs, $s_{\rm ii}$ is either standing (GBGG in
shorthand notation), judging (GBBG), or shunning (GBBB) (refer to
\FIG\ref{fig:norms} for definitions of these norms).  In the
shorthand notation, the first, second, third, and fourth letters
(either G or B) indicate the donor's or $g_{\rm d}$'s new reputation
when the donor cooperates with a G recipient, the donor defects
against a G recipient, the donor cooperates with a B recipient, and
the donor defects against a B recipient, respectively.
Standing, judging, and shunning in $s_{\rm ii}$
are exchangeable for any fixed combination of
$\sigma^{\rm in}=$ Disc, $\sigma^{\rm out}$, $s_{\rm io}$, and $s_{\rm oo}$.
Therefore, there are $270/3=90$ combinations of $\sigma^{\rm out}$,
$s_{\rm io}$, and $s_{\rm oo}$, which are summarized in
\TAB\ref{tab:stable pairs 1}. An asterisk indicates an entry that can be either G or B. For example, GB{\textasteriskcentered}G indicates standing (GBGG) or judging (GBBG).
The probability of
cooperation toward outgroup recipients, payoff ($\pi$; \EQ\eqref{eq:pi}), and the probability that a group has
a G reputation ($p_{\rm g}^*$; \EQ\eqref{eq:pg equilibrium case A}) are also shown in \TAB\ref{tab:stable pairs 1}.
The stable action--norm pairs can be classified into three
categories.

\begin{itemize}
\item Full cooperation: Donors behave as Disc toward outgroup
  recipients, i.e., $\sigma^{\rm out}=$ Disc and cooperate with
  both ingroup and outgroup recipients with probability 1. Accordingly,
$\pi=b-c$ and $p_{\rm g}^*=1$.

In this
  case, indirect reciprocity among different groups as well as that
  within single groups is realized.  Action rule $\sigma^{\rm
    in}=\sigma^{\rm out}=$ Disc is stable if $s_{\rm io}$ is either
  standing (GBGG), judging (GBBG), or shunning (GBBB) and $s_{\rm
    oo}$ is either standing or judging. The condition for stability against
group mutation is the mildest one (i.e., $b>c$) for each action--norm pair.

Under full cooperation, $s_{\rm io}$ and $s_{\rm io}$
must be one that stabilizes cooperation in the
standard indirect reciprocity game without a group-structured population
\citep{OhtsukiIwasa2004jtb,Nowak2005Nature,OhtsukiIwasa2007jtb}.  The
ingroup observer monitors
donors' actions toward outgroup recipients through the use of
$s_{\rm io}=$ standing, judging, or shunning,
even though ingroup players are not directly harmed if
donors defect against outgroup recipients.  The ingroup observer does
so because donors' defection against outgroup recipients would
negatively affect the group's reputation.

\item Partial ingroup favoritism: Donors adopt $\sigma^{\rm out}=$ Disc and
cooperate with ingroup recipients with probability 1 and 
outgroup recipients with probability $1/2$.
Accordingly, $\pi=(b-c)(1+r^{\rm in})/2$ and $p_{\rm g}^*=1/2$.

In this case, action rule $\sigma^{\rm
    in}=\sigma^{\rm out}=$ Disc is stable if
$s_{\rm io}$ is either standing (GBGG) or judging (GBBG),
and $s_{\rm oo}$ is either scoring (GBGB) or shunning (GBBB).
The condition for stability against group mutation
is shown in \TAB\ref{tab:condition partial ingroup favoritism}.

\item Perfect ingroup favoritism: Donors adopt 
$\sigma^{\rm out}=$ AllD and
  always cooperate with ingroup recipients and never with outgroup
  recipients regardless of the recipient's group
  reputation. Accordingly, $\pi=(b-c)r^{\rm in}$.

Table~\ref{tab:stable pairs 1} suggests that
action rule ($\sigma^{\rm in}$, $\sigma^{\rm out}$) $=$ (Disc,
AllD) can be stable for any subnorm $s_{\rm oo}$.
This is true because the group reputation, whose update rule is given by $s_{\rm oo}$, is irrelevant in the current situation; the donor anyways defects against outgroup recipients.  Nevertheless, $s_{\rm oo}$ determines $s_{\rm io}$ that is consistent with ingroup cooperation through the probability of a G group reputation $p_{\rm g}^*$.

When $s_{\rm oo}=$ {\textasteriskcentered}G{\textasteriskcentered}G, the outgroup observer evaluates defection against outgroup
recipients to be G (\FIG\ref{fig:ingroup outgroup observers}B).
Therefore, $p_{\rm g}^*=1$. In this case, $s_{\rm io}=$ {\textasteriskcentered}GBB, {\textasteriskcentered}GBG, and {\textasteriskcentered}GGG stabilize
perfect ingroup favoritism.  Under any of these $s_{\rm io}$,
the ingroup observer assigns G to a donor that defects
against a recipient in a G outgroup because the
second entry of $s_{\rm io}$ is equal to G in each case.
Therefore, $p^*=1$, and full ingroup cooperation is stable.

When $s_{\rm oo}=$ {\textasteriskcentered}G{\textasteriskcentered}B or {\textasteriskcentered}B{\textasteriskcentered}G,
the outgroup observer evaluates defection against outgroup
recipients to be G with probability $1/2$. Therefore, $p_{\rm g}^*=1/2$.
In this case, $s_{\rm io}=$ {\textasteriskcentered}G{\textasteriskcentered}G stabilizes perfect ingroup favoritism.
Under such an $s_{\rm io}$,
the ingroup observer assigns G to a donor that defects against
a recipient in a G outgroup because the second and fourth
entries of $s_{\rm io}$ are equal to G.

When $s_{\rm oo}=$ {\textasteriskcentered}B{\textasteriskcentered}B, the outgroup observer evaluates
defection against outgroup
recipients to be B. Therefore, $p_{\rm g}^*=0$.
In this case, $s_{\rm io}=$ BB{\textasteriskcentered}G, BG{\textasteriskcentered}G, and GG{\textasteriskcentered}G stabilize perfect ingroup favoritism.
Under such an $s_{\rm io}$, the ingroup observer assigns G to a
donor that defects against a recipient in a G outgroup because
the fourth entry of $s_{\rm io}$ is equal to G.

In all the cases, the stability against invasion by group mutants requires
$b>c$. 

\end{itemize}

\subsubsection{Scenario 2}\label{sub:scenario 2}

In scenario 2 of group mutation, it is hypothesized that
a group of mutants immigrates from a different
population that is stable against invasion by single mutants.  Such a
group mutant may appear owing to the encounter of different
stable cultures (i.e., action--norm pairs). The pairs that 
are stable against invasion by single mutants and yield zero payoff,
such as the population of AllD players, must be
also included in the group mutant list.
It should be noted that a mutant group may have a different social norm from that for the resident population.

Among the 588 action--norm pairs that are stable against single mutation,
no pair is stable against group mutation. However,
140 pairs are stable against group mutation for any $b>c$
in a relaxed sense that the resident player's payoff is not smaller
than the group mutant's payoff, i.e., $\pi\ge \pi_{\rm g}^{\prime}$
(\FIG\ref{fig:equilibrium selection}).
The homogeneous population of each pair is neutrally 
invaded by some group mutants, i.e., $\pi=\pi_{\rm g}^{\prime}$.
Therefore, I examine the
evolutionary stability (e.g., \cite{Nowak2006book})
against group mutation. In other words, 
for the group mutants yielding $\pi=\pi_{\rm g}^{\prime}$,
I require
$\pi>\pi_{\rm g}^{\prime}$ when the resident players are replaced by group mutants.

All 140 action--norm pairs are evolutionarily
stable except that each pair is still neutrally invaded by their
cousins. For example, four action--norm pairs specified by $\sigma^{\rm
  in}=\sigma^{\rm out}=$ Disc, $s_{\rm ii}=$ GB{\textasteriskcentered}G, $s_{\rm io}=$ GB{\textasteriskcentered}G,
$s_{\rm oo}=$ GBGG neutrally invade each other. These pairs yield
the same payoff $\pi=b-c$ and are evolutionarily stable against invasion by the 
other group mutants. Therefore, I conclude that the four pairs
collectively form a set of stable solutions.
Other sets of stable solutions consist of four or eight neutrally
invadable action--norm
pairs that yield the same payoff and differ only in $s_{\rm ii}$ and
$s_{\rm io}$.

All 140 pairs realize
perfect intragroup cooperation such that
the players have G personal reputations and $\sigma^{\rm in}=$ Disc
(\FIG\ref{fig:equilibrium selection}). Subnorm
$s_{\rm   ii}=$ GBGG (i.e., standing) or GBBG (i.e., judging)
is exchangeable for any fixed combination of
$\sigma^{\rm in}=$ Disc, $\sigma^{\rm out}$, $s_{\rm io}$, and $s_{\rm oo}$.
Therefore, there are $140/2=70$ possible combinations of
$\sigma^{\rm out}$, $s_{\rm io}$, and $s_{\rm oo}$, which are listed
in \TAB\ref{tab:stable pairs 2}. The 140 pairs are a subset of the
270 pairs stable under scenario 1. The stable sets of action--norm pairs can be classified
into three categories.
(1) Full cooperation occurs if all the subnorms are
standing or judging. As already mentioned as an example,
under $s_{\rm oo}=$ GBGG, the four action--norm pairs
$(\sigma^{\rm in}, \sigma^{\rm out}, s_{\rm ii}, s_{\rm io}) =$ 
(Disc, Disc, GBGG, GBGG), (Disc, Disc, GBGG, GBBG), (Disc, Disc, GBBG,
GBGG), and (Disc, Disc, GBBG, GBBG) can neutrally invade each
other. Similarly, 
if $s_{\rm oo}=$ GBBG, the same four action--norm pairs constitute a
set realizing stable full cooperation. These two sets of four pairs
are evolutionarily stable against invasion by each other. In total, there are eight
pairs that realize full cooperation.
(2) Partial ingroup favoritism occurs for a set of four
action--norm pairs.
(3) Perfect ingroup favoritism occurs under the same
subnorms $s_{\rm oo}$ as those for scenario 1.
For a fixed $s_{\rm oo}$, the same
eight action--norm pairs $(\sigma^{\rm in}, \sigma^{\rm out}, s_{\rm ii}, s_{\rm
  io}) =$ (Disc, AllD, GB{\textasteriskcentered}G, {\textasteriskcentered}G{\textasteriskcentered}G) yield the same payoff
$\pi=(b-c)r^{\rm in}$,
can neutrally invade each other, and
are evolutionarily stable against the other group mutants.

\subsection{When observers use simpler social norms}\label{sub:simpler
norms}

In fact, players may not differentiate between the three subnorms. 
Players may use a common norm for assessing ingroup donors
irrespective of the location of recipients.
Table~\ref{tab:stable pairs 1} indicates that,
if $s_{\rm ii}=s_{\rm io}$ is imposed for the resident population, but
not for mutants, perfect ingroup favoritism is excluded.
Under scenario 1,
full cooperation is stable
when $s_{\rm ii}=s_{\rm io}=$ standing, judging, or shunning
and $s_{\rm oo}=$ standing or judging. Partial ingroup favoritism
is stable when $s_{\rm ii}=s_{\rm io}=$ standing or judging and
$s_{\rm oo}=$ scoring or shunning. Under scenario 2, full
cooperation is stable when
$s_{\rm ii}=s_{\rm io}=$ standing or judging
and $s_{\rm oo}=$ standing or judging. Partial ingroup favoritism
is stable when $s_{\rm ii}=s_{\rm io}=$ standing or judging and
$s_{\rm oo}=$ shunning.

Alternatively, players may use a common norm for assessing donors
playing with outgroup recipients irrespective of the location of donors.
If $s_{\rm ii}\neq s_{\rm io}$ is allowed and
$s_{\rm io}=s_{\rm oo}$ is imposed, partial ingroup favoritism is excluded.
Under scenario 1,
full cooperation is stable when
$s_{\rm ii}=$ standing, judging, or
shunning and $s_{\rm io}=s_{\rm oo}=$ standing or judging.
Perfect ingroup favoritism is stable when
$s_{\rm ii}=$ standing, judging, or
shunning and $s_{\rm io}=s_{\rm oo}=$ {\textasteriskcentered}G{\textasteriskcentered}G.
The results under scenario 2 differ from
those under scenario 1 only in that $s_{\rm ii}=$ shunning is disallowed.

Finally, if all the three subnorms are forced to be equal, only
full cooperation is stable, and the norm is standing or judging. This holds
true for both scenarios 1 and 2.

\section{Discussion}\label{sec:discussion}

\subsection{Summary of the results}

I identified the pairs of action rule and social norm that are stable
against invasion by single and group mutants in the game of group-structured indirect reciprocity.
Full cooperation (i.e., cooperation within and across groups)
based on personal and group reputations,
partial ingroup favoritism, and perfect ingroup favoritism
are stable under different social norms.
Perfect ingroup favoritism is attained
only when the donor defects against outgroup
recipients regardless of their reputation (i.e., $\sigma^{\rm out}=$
AllD). Perfect ingroup favoritism does not occur with the combination of
a donor that is ready to cooperate with G outgroup recipients
(i.e., $\sigma^{\rm out}=$ Disc) and a B group reputation.
The mechanism for ingroup favoritism revealed in this study 
is distinct from those proposed previously
(see \SEC\ref{sec:introduction}).

The major condition for either full cooperation, partial ingroup favoritism,
and perfect ingroup favoritism, depending on the assumed social norm,
is given by $b r^{\rm in}>c$. In only 3 out of 270 social norms in
scenario 1, an additional condition for $r^{\rm in}$ is imposed
(\SEC\ref{sub:scenario 1}). In general, different mechanisms of cooperation
can be understood in an unified manner such that cooperation occurs if
and only if $b/c$ is larger than a threshold value \citep{Nowak2006Science}.
For example, $b/c$ must be larger than the inverse of the relatedness parameter
$r$ and the inverse of the discount factor in
kin selection and direct reciprocity, respectively. The present result
also fits this view; $r^{\rm in}$ corresponds to $r$ in the case of
kin selection.

I assumed that players approximate personal
reputations of individuals in other groups by group reputations (i.e.,
outgroup homogeneity). Adoption of outgroup homogeneity may be
evolutionarily beneficial for players owing to the reduction in the
cognitive burden of recognizing others' personal reputations.  Instead, the
players pay potential costs of not being able to know the personal
reputations of individuals in other groups. To
explore evolutionary origins of group reputation, one has to
examine competition between players using the group reputation and players not
using it. It would also be necessary to introduce a parameter representing
the cost of obtaining personal reputations of outgroup individuals.
Such an analysis is warranted for future work.

All the players are assumed to use the same social norm. This
  assumption may be justified for well-mixed populations but less so
  for populations with group structure because group structure implies
  relatively little intergroup communication. It seems to be more natural
  to assume that subnorms $s_{\rm ii}$ and $s_{\rm io}$, which are
  used to evaluate actions of ingroup donors, depend on groups.
Under scenario 2 (\SEC\ref{sub:scenario 2}), any stable action--norm
  pair is neutrally invaded by its cousins who are different in
  $s_{\rm ii}$ and $s_{\rm io}$.  This result implies that different groups
  can use different norms. For example, for all the solutions shown
  in Table~\ref{tab:stable pairs 2}, some 
groups can use $s_{\rm ii}=$ GBGG (i.e.,
standing), while other groups in the same population
can use $s_{\rm io}=$ GBBG (i.e.,
  judging). To better understand the possibility of heterogeneous social
  norms, analyzing
a population composed of a small number of groups, probably by
different methods, would be helpful.

\subsection{Cooperation based on group reputation is
distinct from group selection}\label{sub:group selection}

Indirect reciprocity based on group reputation is
distinct from any type of group selection. This is true
for both full cooperation and ingroup favoritism.
There are two dominant variants of group selection that serve as mechanisms for
cooperation in social dilemma games \citep{West2007JEB,West2008JEB}.

The first type is group competition, in which selection pressure
acts on groups such that a group with a large mean payoff would
replace one with a small mean payoff. Models with group competition
induce ingroup favoritism \citep{ChoiBowles2007Science,Garcia2011EHB},
altruistic punishment \citep{Boyd2003PNAS}, and evolution of the
judging social norm in the standard game of indirect reciprocity
whereby players interact within each group
\citep{Pacheco2006PlosComputBiol,Chalub2006jtb}.  In contrast, the
present study is not concerned with evolutionary dynamics
including group competition.  The group mutant is assumed to
statically compare the payoff to the resident group with that to the
mutant group.

The second type of group selection requires assortative reproduction
in the sense that the offspring have a higher probability of belonging to
specific groups than to other groups depending on the offspring's
genotype. It is mathematically identical with kin selection \citep{West2007JEB,West2008JEB}. This variant of group selection is also irrelevant to the
present model, which is not concerned with the reproduction process.

The analysis in this study is purely static.
I avoided examining evolutionary dynamics for two
reasons. First, the discovered mechanism for cooperation may
be confused with group selection in the presence of evolutionary dynamics.
Second, the model becomes
needlessly complicated.
Introducing evolutionary dynamics implies that one specifies a
rule for reproduction. 
Offspring may be assumed to belong to the
parent's group or to migrate to another group. It may then be
necessary to consider the treatment of, for example, the heterogeneous
group size.
Because evolutionary dynamics are neglected, the present model
explains neither emergence of 
full cooperation and ingroup favoritism nor
the likelihood of different solutions, which is a
main limitation of the present study.

I stress that the concept of group mutants is introduced
to sift the set of stable action--norm pairs. Unless group
competition is assumed, the concept of group mutants
does not particularly promote cooperation in evolutionary dynamics.

\subsection{Group competition can enable full cooperation and ingroup
  favoritism even if $b r^{\rm in}>c$ is violated}

Under a proper social norm, full cooperation or ingroup favoritism
is stable if $b r^{\rm in}>c$ (i.e., \EQ\eqref{eq:b rin > c} is satisfied)
in most cases.
With probability $r^{\rm
  in}$, the donor, recipient, and observer are
engaged in the standard (i.e., no group structure) indirect
reciprocity game limited to a single group (\FIG\ref{fig:ingroup outgroup observers}A).
In the standard indirect
reciprocity game under incomplete information, $bq>c$ is quite often
the condition for cooperation, where $q$ is the
probability that the recipient's reputation is observed. This holds
true when $q$ indicates the observation probability for the donor
\citep{Nowak1998nature,Nowak1998jtb,Brandt2005PNAS,Brandt2006jtb,SuzukiToquenaga2005JTB}
or that for both the donor and observer
\citep{NakamuraMasuda2011PLoSComputBiol}. Because $r^{\rm in}$ is also
equal to the probability that the donor sees the recipient's
personal reputation, $r^{\rm in}$ resembles $q$. In fact, replacing $r^{\rm in}$ by
$q$ in \EQ\eqref{eq:b rin > c} yields $bq>c$.

If a player is capable of recognizing the personal reputation of a
fixed number of others, the maximum population
size for which indirect reciprocity is possible in the standard
indirect reciprocity game scales as $1/q$. The consistency between
\EQ\eqref{eq:b rin > c} and $bq>c$ implies that the concept of group
reputation does not increase the maximum population size for which indirect reciprocity occurs.
However, under group competition
(\SEC\ref{sub:group selection}), 
full cooperation and ingroup favoritism can be stable even if
the restriction imposed by \EQ\eqref{eq:b rin > c} is removed.

To explain this point, assume that the population is subjected to
evolutionary dynamics such that players with relatively large payoffs
would bear more offspring in the same group and group competition occurs.
The rate of group competition is denoted by
$1/t_{\rm gc}$, where $t_{\rm gc}$ is the mean time interval between
successive group competition events. Emergence of a single mutant
occurs with rate $1/t_{\rm m}$. Selection and reproduction of single
players occur with rate $1/t_{\rm s}$.

If \EQ\eqref{eq:b rin > c} is violated, single mutants emerge in time
$\propto t_{\rm m}$. Then, some types of mutants, including the AllD
mutant, spread in the invaded group in
time $\propto t_{\rm s}$ under scenario 1 of group mutation.
The invaded group presumably possesses a
smaller group-averaged payoff than other resident groups
because the resident population is stable against
invasion by group mutants as long as $b>c$, in all but three of 270
action--norm pairs (\TAB\ref{tab:condition partial ingroup favoritism}).
If $1/t_{\rm gc}\gg 1/t_{\rm m}$, 
such an invaded group is likely to be eradicated by group competition
because group competition
occurs much faster than the emergence of single mutants.
In this case, full cooperation or ingroup favoritism, depending on the
given social norm, can be maintained in the absence of
\EQ\eqref{eq:b rin > c}. This discussion
does not involve timescale $t_{\rm s}$.

Group competition is needed to remove \EQ\eqref{eq:b rin
  > c}.  If \EQ\eqref{eq:b rin > c} is imposed, cooperation occurs
without group competition.

\subsection{Relationship to previous behavioral experiments}

In this section, I discuss possible linkages between the present model and the previous experiments examining indirect reciprocity and third-party punishments.

Yamagishi and colleagues conducted a series of laboratory experiments to show that
ingroup favoritism is induced by a group heuristic
\citep{Yamagishi1998AJSP,Yamagishi1999chapter,YamagishiMifune2008RS}. With
a group heuristic,
donors cooperate with ingroup recipients because the donors expect
repayment from other ingroup players. Donors do not use the information about others' reputations in these experiments. In contrast,
players use personal reputations of ingroup members
in the present model. Nevertheless, the previous experiments and the current model do
not contradict each other. 

In another laboratory experiment, Mifune et al. showed that
presentation of eye-like painting
promotes donor's cooperation toward ingroup recipients in the dictator
game \citep{Mifune2010EHB}.
For expository purposes,
I define serious subnorm to be either standing,
judging, or shunning.
If the eye-like painting
approximates an ingroup observer obeying a serious subnorm,
this experimental result is consistent with the present theory
because ingroup cooperation is theoretically stable when
the ingroup observer adopts a serious subnorm.
Because the painting does not 
increase the cooperation toward outgroup recipients
\citep{Mifune2010EHB}, it may not turn $s_{\rm io}$ to a serious subnorm
for some psychological reason.
Humans may use double standards, i.e., $s_{\rm ii}\neq s_{\rm io}$,
which favor ingroup favoritism in my model.

Other behavioral experiments have addressed the relationship between
third-party altruistic punishments and ingroup favoritism
\citep{Bernhard2006Nature,Shinada2004EHB}. In precise terms,
third-party punishments and reputation-based indirect reciprocity are
distinct mechanisms for cooperation
\citep{Sigmund2001PNAS,Ohtsuki2009Nature}. Nevertheless, below I discuss
possible linkages between these experiments and my model.

In indigenous
communities in Papua New Guinea \citep{Bernhard2006Nature}, the amount of punishment is larger if
the punisher belongs to the donor's group than to a different group
(compare ABC and AB cases in their Fig. 1).
Their results
suggest that the ingroup observer may use a serious subnorm
and the outgroup observer may not.
Furthermore, given
that the punisher is in the donor's group,
the amount of punishment is larger if the donor and recipient belong to
the same group (\FIG\ref{fig:ingroup outgroup observers}A, if
the punisher is identified with the ingroup observer) than if they belong to
different groups (\FIG\ref{fig:ingroup outgroup observers}B;
compare the ABC and AC cases in Fig. 1 of \cite{Bernhard2006Nature}). In this situation,
the ingroup observer may use a serious subnorm $s_{\rm ii}$
when the donor plays with ingroup recipients (\FIG\ref{fig:ingroup
  outgroup observers}A) and use a nonserious subnorm $s_{\rm io}$ when
the donor plays with outgroup recipients (\FIG\ref{fig:ingroup
  outgroup observers}B).
My model reproduces ingroup favoritism under these conditions.

However, my model and others
are not concerned with
a main finding in \citep{Bernhard2006Nature} that
the amount of punishment is larger when the punisher and recipient
belong to the same group.
For the reasons stated in \SEC\ref{sub:update},
I did not assume that observers make their judgments differently
when they belong to the recipient's group $g_{\rm r}$ and to a different
group. To theoretically explain
the main finding in \cite{Bernhard2006Nature},
one should explicitly analyze the case of a finite number of groups.

In different laboratory experiments,
the amount of punishment is larger
for an ingroup donor's defection than an outgroup donor's defection
\citep{Shinada2004EHB}.
My results are consistent with their results
in that, for ingroup favoritism,
the donor's action must be seriously evaluated
by the ingroup observer using $s_{\rm ii}$ and not seriously
by the outgroup observer using $s_{\rm oo}$.

\subsection{Reduction of ingroup favoritism}

Although ingroup favoritism seems to be a canonical behavior of humans,
reduction of ingroup bias would induce intergroup cooperation and is
socially preferable \citep{Yamagishi1998AJSP}.
Full cooperation is Pareto efficient, whereas ingroup favoritism is not.
Various psychological and sociological mechanisms
for reducing the ingroup bias, such as
guilt, ``auto-motive'' control, retraining, empathy, 
and decategorization have been proposed \citep{Hewstone2002ARP,Dovidio2005book,Sedikides1998book}.

My results provide theory-based possibilities of reducing ingroup bias.
First, if the social norm is fixed,
conversion from ingroup favoritism to full cooperation is theoretically
impossible because
full cooperation and ingroup favoritism do not
coexist under a given social norm.
Therefore, advising players to
change their behavior toward outgroup recipients
from AllD to Disc is not recommended unless the social norm is also altered.
Conversion from ingroup favoritism to
full cooperation requires a change in the social
norm such that players as observers seriously assess
ingroup donors' actions toward outgroup recipients (with $s_{\rm
io}$) and outgroup--outgroup interaction (with $s_{\rm oo}$).
In particular, if $s_{\rm io}$ is a serious subnorm, perfect ingroup
favoritism with no intergroup cooperation disappears
(\SEC\ref{sub:simpler norms}).

Second, if the three subnorms are the same, the perfect and partial ingroup
favoritism is eradicated. 
The coincidence of only two subnorms is insufficient to induce
full cooperation (\SEC\ref{sub:simpler norms}).
The subnorms $s_{\rm ii}=s_{\rm io}=s_{\rm oo}$ that exclude the ingroup bias and realize full
cooperation are standing or judging. Therefore, without speaking of
serious subnorms, forcing players to
use the same subnorms consistently in assessing donors in different situations
may be also effective in inducing full cooperation. 

Ingroup favoritism has been mostly an experimental question except for
some recent theoretical studies. This study is a first step toward understanding and
even manipulating the dichotomy between full cooperation and ingroup
favoritism in the context of indirect reciprocity.

\newpage

\setcounter{section}{0}
\renewcommand{\thesection}{\Alph{section}}

\newcounter{appendixcount}
\renewcommand{\theappendixcount}{\Alph{appendixcount}}
\newcommand{\appendixcount}{\refstepcounter{appendixcount}}

\section*{Appendix A: A variant of the model with different reputation dynamics}\appendixcount\label{sec:appendix variant}

In this section, I analyze a variant of the model in which outgroup
observers update the group reputation of donors involved in ingroup
interaction (i.e., $g_{\rm d}=g_{\rm r}$). 

\subsection*{Reputation dynamics}

I assume that the outgroup observer uses the donor's action, the 
recipient's personal reputation, and $s_{\rm oo}$, to update $g_{\rm
  d}$'s (not the donor's personal) reputation.

The equivalent of \EQ\eqref{eq:pg equilibrium case A} under this
reputation update rule is given by
\begin{equation}
p_{\rm g}^* = r^{\rm in}\left[ p^*\Phi_{\rm G}^{\rm out}(\sigma^{\rm in})
+ (1-p^*)\Phi_{\rm B}^{\rm out}(\sigma^{\rm in})\right]
 + r^{\rm out} \left[
p_{\rm g}^* \Phi_{\rm G}^{\rm out}(\sigma^{\rm out})
+ (1-p_{\rm g}^*) \Phi_{\rm B}^{\rm out}(\sigma^{\rm out})
 \right].
\label{eq:p_g equilibrium case B}
\end{equation}
I obtain $p^*$ and $p_{\rm g}^*$ by solving the set of linear
equations \eqref{eq:p equilibrium} and \eqref{eq:p_g equilibrium case B}.
Equations~\eqref{eq:p'*}--\eqref{eq:p'* group mutation}, and
\eqref{eq: p'* group mutation final}
are unchanged. As compared to the case of the original reputation
update rule (original case for short), \EQ~\eqref{eq:pg'* group mutation case A}
is replaced by
\begin{align}
p_{\rm g}^{\prime *} =& r^{\rm in} 
\left[ p^{\prime *}\Phi_{\rm G}^{\rm out}(\sigma^{\rm in\prime}) + (1-p^{\prime
  *}) \Phi_{\rm B}^{\rm out}(\sigma^{\rm in\prime})\right]\notag\\
&+ r^{\rm out}\left[
p_{\rm g}^*\Phi_{\rm G}^{\rm out}(\sigma^{\rm out\prime}) + 
(1-p_{\rm g}^*)\Phi_{\rm B}^{\rm out}(\sigma^{\rm out\prime})
\right].\label{eq:pg'* group mutation case B}
\end{align}
The equivalent of \EQ\eqref{eq: pg'* group mutation final case A} is
obtained by substituting
\EQ\eqref{eq: p'* group mutation final} in \EQ\eqref{eq:pg'* group
mutation case B}.

Because of
the symmetry with respect to G and B, I exclude action rules having
$\sigma^{\rm in}=$ AntiDisc from the exhaustive search, as I did in the
original case (\SEC\ref{sub:against single mutants}). 
It should be
noted that one cannot eliminate action--norm pairs with
$\sigma^{\rm out}=$ AntiDisc
on the basis of symmetry consideration, which is different from the original case.
This is because a player's personal and group reputations are
interrelated through the behavior of the outgroup observer when
$g_{\rm d}=g_{\rm r}$.

\subsection*{Results}

Under the modified reputation update rule, there are
725 action--norm pairs that are stable against invasion by single mutants
and yield $\pi>0$.

Under scenario 1, 507 out of the 725 pairs are stable against group
mutation, and 324 out of the 507 pairs yield perfect ingroup cooperation. 
The 324 action--norm pairs are classified as follows.
First, 68 pairs yield full cooperation with either $(\sigma^{\rm in},
\sigma^{\rm out})=$ (Disc, Disc) or (Disc, AntiDisc).
Second, 14 pairs yield partial ingroup favoritism with
$(\sigma^{\rm in}, \sigma^{\rm out})=$ (Disc, AntiDisc).
Third, 236 pairs yield perfect ingroup favoritism with
$(\sigma^{\rm in}, \sigma^{\rm out})=$ (Disc, AllD).
Fourth, 6 pairs yield perfect ingroup favoritism with
$(\sigma^{\rm in}, \sigma^{\rm out})=$ (Disc, AntiDisc).

As in the original case,
$\sigma^{\rm in}=$ Disc, and
$s_{\rm ii}$ is either standing,
judging, or shunning for these pairs.
In contrast to the original case,
$(\sigma^{\rm in}, \sigma^{\rm out})=$ (Disc, AntiDisc) can be stable,
yield perfect ingroup cooperation, and even yield
outgroup cooperation, under some social norms.
In such a situation,
the values of the personal and group reputations (i.e., G and B)
have opposite meanings.
In other words, a G but not B personal reputation elicits intragroup cooperation,
while a B but not G group reputation elicits intergroup cooperation.
Therefore, action rule $(\sigma^{\rm in}, \sigma^{\rm out})=$ (Disc, AntiDisc) in this situation can be regarded as a relative of 
$(\sigma^{\rm in}, \sigma^{\rm out})=$ (Disc, Disc) in the situation in which the values of the personal and group reputations have the same meaning.
On this basis, I consider that the present results are similar to
those obtained for the original case (\TAB\ref{tab:stable pairs 1}).
In particular, only full cooperation is stable under standing or judging
if $s_{\rm ii}$, $s_{\rm io}$, and $s_{\rm oo}$
are assumed to be the same.

Under scenario 2,
144 out of 725 pairs are stable against group mutation, and all of them yield
perfect ingroup cooperation. The 
140 pairs that survive in the original case
(\SEC\ref{sub:scenario 2}) also survive under the modified reputation
update rule.
The action rule in the additional four ($=144-140$) pairs is
$(\sigma^{\rm in}, \sigma^{\rm out})=$ (Disc, AntiDisc).
Another
difference from the original case
is that the action--norm pairs that yield partial ingroup
favoritism in \TAB\ref{tab:stable pairs 2} realize full cooperation in
the present case. Otherwise, the results are the same as those in the
original case.
In summary, 16 pairs realize full cooperation, and
128 pairs realize perfect ingroup favoritism.
As is the case for scenario 1,
only full cooperation is stable with standing or judging
if the three subnorms are assumed to be the same.

\section*{Appendix B: The rest of the stable action--norm pairs under scenario 1}\appendixcount\label{sec:outgroup favoritism}

Under sceinario 1 in the original case, 270 out of 440 stable action--norm pairs with a positive payoff realize perfect intragroup cooperation (\SEC\ref{sub:scenario 1}).
The other 170 stable action--norm pairs yielding $\pi>0$ are
summarized in \TAB\ref{tab:stable pairs 1 imperfect}. 
For all the stable action--norm pairs shown, $\sigma^{\rm in}=$ Disc. Table~\ref{tab:stable pairs 1 imperfect} indicates that
outgroup favoritism does not occur.

There are 18 rows in \TAB\ref{tab:stable pairs 1 imperfect}.
For the two action--norm pairs shown in the first row, the stability condition
is given by $br^{\rm in}>c$ and $r^{\rm in}<1/2$. For the two action--norm pairs shown in the sixth row, the stability condition is given by
$br^{\rm in}>c$ and $r^{\rm in}>\sqrt{2}-1$. For the four action--norm pairs shown in the sixteenth row, the stability condition is given by $b/c>(1+r^{\rm in})/r^{\rm in}$. For all the other action--norm pairs, the stability condition is given by $br^{\rm in}>c$.

\section*{Acknowledgements}

I thank Mitsuhiro Nakamura and Hisashi Ohtsuki for valuable discussions and 
acknowledge the support provided through Grants-in-Aid for Scientific Research (Nos. 20760258 and 23681033, and Innovative Areas ``Systems Molecular Ethology''(No. 20115009)) from MEXT, Japan.

\newpage
\clearpage

\clearpage
\clearpage

\clearpage

\pagestyle{empty}

\begin{table}
\begin{center}
\caption{Stable action--norm pairs with perfect ingroup cooperation
under scenario 1.
The probability of cooperation with outgroup recipients, $\pi$, and $p_{\rm
  g}^*$ are the values in the limit $\epsilon\to 0$. 
$s_{\rm ii}=$ GBGG (standing), GBBG (judging), or GBBB (shunning).
Action--norm pairs only different in
$s_{\rm ii}$ were distinguished when counting the number of stable action--norm pairs. An asterisk indicates that both G and B apply.}
\label{tab:stable pairs 1}
\begin{tabular}{|c|c|c|c|c|c|c|}\hline
\multirow{2}{*}{State} & Prob. C to & \multirow{2}{*}{$\pi$} & \multirow{2}{*}{$\sigma^{\rm out}$} & \multirow{2}{*}{$p_{\rm g}^*$} & Social norm 
 & No.\\
& outgroup &&&& ($s_{\rm io}$--$s_{\rm oo}$) &  pairs \\ \hline
\multirow{2}{*}{Full cooperation} & \multirow{2}{*}{1} & \multirow{2}{*}{$b-c$} & \multirow{2}{*}{Disc} & \multirow{2}{*}{1} & GB{\textasteriskcentered}G--GB{\textasteriskcentered}G & \multirow{2}{*}{18}\\
 &&&&& GBBB--GB{\textasteriskcentered}G &\\ \hline
Partial ingroup & \multirow{2}{*}{$\frac{1}{2}$} & \multirow{2}{*}{$\frac{(b-c)(1+r^{\rm in})}{2}$} & \multirow{2}{*}{Disc} & \multirow{2}{*}{$\frac{1}{2}$} & \multirow{2}{*}{GB{\textasteriskcentered}G-GB{\textasteriskcentered}B} & \multirow{2}{*}{12}\\
 favoritism &&&&&& \\ \hline
 & \multirow{8}{*}{0} & \multirow{8}{*}{$(b-c)r^{\rm in}$} & \multirow{8}{*}{AllD} & \multirow{3}{*}{1} & {\textasteriskcentered}GBB--{\textasteriskcentered}G{\textasteriskcentered}G & \multirow{3}{*}{72}\\
  &&&&& {\textasteriskcentered}GBG--{\textasteriskcentered}G{\textasteriskcentered}G & \\
 &&&&& {\textasteriskcentered}GGG--{\textasteriskcentered}G{\textasteriskcentered}G & \\ \cline{5-7}
Perfect ingroup  &&&& \multirow{2}{*}{$\frac{1}{2}$} & {\textasteriskcentered}G{\textasteriskcentered}G--{\textasteriskcentered}G{\textasteriskcentered}B & \multirow{2}{*}{96}\\
 favoritism &&&&& {\textasteriskcentered}G{\textasteriskcentered}G--{\textasteriskcentered}B{\textasteriskcentered}G & \\ \cline{5-7}
 &&&& \multirow{3}{*}{0} & BB{\textasteriskcentered}G--{\textasteriskcentered}B{\textasteriskcentered}B & \multirow{3}{*}{72}\\
 &&&&& BG{\textasteriskcentered}G--{\textasteriskcentered}B{\textasteriskcentered}B & \\
&&&&& GG{\textasteriskcentered}G--{\textasteriskcentered}B{\textasteriskcentered}B & \\ \hline
\end{tabular}
\end{center}
\end{table}

\newpage
\clearpage

\begin{table}
\begin{center}
\caption{Conditions for stability of partial ingroup favoritism
against group mutation under scenario 1.
The condition on $r^{\rm in}$ is required for the three out of 12
social norms
to prevent the invasion by group
mutants that
defect against ingroup recipients and cooperate with outgroup recipients.
}
\label{tab:condition partial ingroup favoritism}
\begin{tabular}{|c|c|c|c|}\hline
\multirow{2}{*}{Conditions} & Social norm & Social norm & No.\\
& ($s_{\rm ii}$) & ($s_{\rm io}$--$s_{\rm oo}$) &  pairs \\ \hline
\multirow{3}{*}{$b>c$} & \multirow{3}{*}{GBGG, GBBG, or GBBB} & GBBG--GBBB
& \multirow{3}{*}{9}\\
&& GBGG--GBBB & \\
&& GBBG--GBGB & \\ \hline
$b>c$ and $r^{\rm in}>\sqrt{2}-1$ & GBGG & GBGG--GBGB & 1\\ \hline
$b>c$ and $r^{\rm in}>1/2$ & GBBG or GBBB & GBGG--GBGB & 2\\ \hline
\end{tabular}
\end{center}
\end{table}

\newpage
\clearpage

%
%
%
%
%
%
\begin{table}
\begin{center}
\caption{Stable action--norm pairs with perfect ingroup cooperation
under scenario 2. $s_{\rm ii}=$ GBGG
(standing) or GBBG (judging).
Different action--norm pairs in the same row are neutrally invadable to each other. An asterisk indicates either G or B.}
\label{tab:stable pairs 2}
\begin{tabular}{|c|c|c|c|c|c|c|}\hline
\multirow{2}{*}{State} & Prob. C to & \multirow{2}{*}{$\pi$} & \multirow{2}{*}{$\sigma^{\rm out}$} & \multirow{2}{*}{$p_{\rm g}^*$} & Social norm
 & No.\\
& outgroup &&&& ($s_{\rm io}$--$s_{\rm oo}$) &  pairs \\ \hline
\multirow{2}{*}{Full cooperation} & \multirow{2}{*}{1} & \multirow{2}{*}{$b-c$} & \multirow{2}{*}{Disc} & \multirow{2}{*}{1} & GB{\textasteriskcentered}G--GBGG & \multirow{2}{*}{8}\\
 &&&&& GB{\textasteriskcentered}G--GBBG &\\ \hline
Partial ingroup & \multirow{2}{*}{$\frac{1}{2}$} & \multirow{2}{*}{$\frac{(b-c)(1+r^{\rm in})}{2}$} & \multirow{2}{*}{Disc} & \multirow{2}{*}{$\frac{1}{2}$} & \multirow{2}{*}{GB{\textasteriskcentered}G--GBBB} & \multirow{2}{*}{4}\\
favoritism &&&&& & \\ \hline
 & \multirow{16}{*}{0} & \multirow{16}{*}{$(b-c)r^{\rm in}$} & \multirow{16}{*}{AllD} & \multirow{4}{*}{1} & {\textasteriskcentered}G{\textasteriskcentered}G--BGBG & \multirow{4}{*}{32}\\
 &&&&& {\textasteriskcentered}G{\textasteriskcentered}G--GGBG & \\
 &&&&& {\textasteriskcentered}G{\textasteriskcentered}G--BGGG & \\
 &&&&& {\textasteriskcentered}G{\textasteriskcentered}G--GGGG & \\ \cline{5-7}
  &&&& \multirow{8}{*}{$\frac{1}{2}$} & {\textasteriskcentered}G{\textasteriskcentered}G--BGBB & \multirow{8}{*}{64}\\
 &&&&& {\textasteriskcentered}G{\textasteriskcentered}G--GGBB & \\
 &&&&& {\textasteriskcentered}G{\textasteriskcentered}G--BGGB & \\
Perfect ingroup &&&&& {\textasteriskcentered}G{\textasteriskcentered}G--GGGB & \\
 favoritism &&&&& {\textasteriskcentered}G{\textasteriskcentered}G--BBBG & \\
 &&&&& {\textasteriskcentered}G{\textasteriskcentered}G--GBBG & \\
 &&&&& {\textasteriskcentered}G{\textasteriskcentered}G--BBGG & \\
 &&&&& {\textasteriskcentered}G{\textasteriskcentered}G--GBGG & \\ \cline{5-7}
 &&&& \multirow{4}{*}{0} & {\textasteriskcentered}G{\textasteriskcentered}G--BBBB & \multirow{4}{*}{32}\\
 &&&&& {\textasteriskcentered}G{\textasteriskcentered}G--GBBB & \\
 &&&&& {\textasteriskcentered}G{\textasteriskcentered}G--BBGB & \\
 &&&&& {\textasteriskcentered}G{\textasteriskcentered}G--GBGB & \\ \hline
\end{tabular}
\end{center}
\end{table}

\newpage

\begin{table}
\begin{center}
\caption{Stable action--norm pairs with a positive probability of cooperation that are not included in \TAB\ref{tab:stable pairs 1}.
$s_{\rm ii}=$ GBGG (standing), GBBG (judging), or GBBB (shunning).
An asterisk indicates either G or B. The sixth and seventh rows in the table are not aggregated because the stability condition is different between these cases (Appendix~\ref{sec:outgroup favoritism}).}
\label{tab:stable pairs 1 imperfect}
\begin{tabular}{|c|c|c|c|c|c|c|c|}\hline
Prob. C to & Prob. C to & \multirow{2}{*}{$\pi$} & \multirow{2}{*}{$\sigma^{\rm out}$} & \multirow{2}{*}{$p^*$} & \multirow{2}{*}{$p_{\rm g}^*$} & Social norm 
 & No.\\
ingroup & outgroup &&&&& ($s_{\rm ii}$--$s_{\rm io}$--$s_{\rm oo}$) &  pairs \\ \hline
$\frac{1}{2}$ & $\frac{1}{2}$ & $\frac{b-c}{2}$ & Disc & $\frac{1}{2}$ & $\frac{1}{2}$ & GBBB-GBBB--GB{\textasteriskcentered}B & 2\\ \hline
\multirow{4}{*}{$\frac{1}{2}$} & \multirow{4}{*}{0} & \multirow{4}{*}{$\frac{(b-c)r^{\rm in}}{2}$} & \multirow{4}{*}{AllD} & \multirow{4}{*}{$\frac{1}{2}$} & \multirow{4}{*}{$\frac{1}{2}$} & {GBBB--{\textasteriskcentered}GBB-{\textasteriskcentered}G{\textasteriskcentered}B} & \multirow{4}{*}{32}\\
&&&&&& GBBB--BB{\textasteriskcentered}G-{\textasteriskcentered}G{\textasteriskcentered}B & \\
&&&&&& GBBB--{\textasteriskcentered}GBB-{\textasteriskcentered}B{\textasteriskcentered}G & \\
&&&&&& GBBB--BB{\textasteriskcentered}G-{\textasteriskcentered}B{\textasteriskcentered}G & \\ \hline
\multirow{2}{*}{$\frac{1+r^{\rm in}}{2}$} & \multirow{2}{*}{$\frac{1}{2}$} & \multirow{2}{*}{$\frac{(b-c)(1+(r^{\rm in})^2)}{2}$} & \multirow{2}{*}{Disc} & \multirow{2}{*}{$\frac{1+r^{\rm in}}{2}$} & \multirow{2}{*}{$\frac{1}{2}$} & GB{\textasteriskcentered}G--GBBB-GBGB & \multirow{2}{*}{4}\\
&&&&&& GB{\textasteriskcentered}G--GBBB-GBBB & \\ \hline
\multirow{4}{*}{$\frac{1+r^{\rm in}}{2}$} & \multirow{4}{*}{0} & \multirow{4}{*}{$\frac{(b-c)r^{\rm in}(1+r^{\rm in})}{2}$} & \multirow{4}{*}{AllD} & \multirow{4}{*}{$\frac{1+r^{\rm in}}{2}$} & \multirow{4}{*}{$\frac{1}{2}$} & GB{\textasteriskcentered}G--{\textasteriskcentered}GBB--{\textasteriskcentered}G{\textasteriskcentered}B & \multirow{4}{*}{64}\\
&&&&&& GB{\textasteriskcentered}G--BB{\textasteriskcentered}G--{\textasteriskcentered}G{\textasteriskcentered}B & \\
&&&&&& GB{\textasteriskcentered}G--{\textasteriskcentered}GBB--{\textasteriskcentered}B{\textasteriskcentered}G & \\
&&&&&& GB{\textasteriskcentered}G--BB{\textasteriskcentered}G--{\textasteriskcentered}B{\textasteriskcentered}G & \\ \hline
\multirow{7}{*}{$r^{\rm in}$} & \multirow{7}{*}{0} & \multirow{7}{*}{$(b-c)(r^{\rm in})^2$} & \multirow{4}{*}{AllD} & \multirow{7}{*}{$r^{\rm in}$} & \multirow{2}{*}{1} & GB{\textasteriskcentered}G--BBBB-{\textasteriskcentered}G{\textasteriskcentered}G & \multirow{7}{*}{68} \\
&&&&&& GB{\textasteriskcentered}G--BB{\textasteriskcentered}G--{\textasteriskcentered}G{\textasteriskcentered}G & \\ \cline{6-7}
&&&&& \multirow{2}{*}{$\frac{1}{2}$} & GB{\textasteriskcentered}G--BBBB--{\textasteriskcentered}G{\textasteriskcentered}B & \\
&&&&&& GB{\textasteriskcentered}G--BBBB--{\textasteriskcentered}B{\textasteriskcentered}G & \\ \cline{4-4} \cline{6-7}
&&& Disc && \multirow{3}{*}{0} & GB{\textasteriskcentered}G--GBBB--BB{\textasteriskcentered}B & \\ \cline{4-4}
&&& \multirow{2}{*}{AllD} &&& GB{\textasteriskcentered}G--BBBB--{\textasteriskcentered}B{\textasteriskcentered}B & \\
&&&&&& GB{\textasteriskcentered}G--{\textasteriskcentered}GBB--{\textasteriskcentered}B{\textasteriskcentered}B & \\ \hline
\end{tabular}
\end{center}
\end{table}


\begin{thebibliography}{50}
\expandafter\ifx\csname natexlab\endcsname\relax\def\natexlab#1{#1}\fi
\expandafter\ifx\csname url\endcsname\relax
  \def\url#1{\texttt{#1}}\fi
\expandafter\ifx\csname urlprefix\endcsname\relax\def\urlprefix{URL }\fi

\bibitem[{Antal et~al.(2009)Antal, Ohtsuki, Wakeley, Taylor, and
  Nowak}]{Antal2009PNAS}
Antal, T., Ohtsuki, H., Wakeley, J., Taylor, P.~D., Nowak, M.~A., 2009.
  {Evolution of cooperation by phenotypic similarity}. Proc. Natl. Acad. Sci.
  USA 106, 8597--8600.

\bibitem[{Axelrod(1984)}]{Axelrod1984book}
Axelrod, R., 1984. Evolution of Cooperation. Basic Books, NY.

\bibitem[{Bernhard et~al.(2006)Bernhard, Fischbacher, and
  Fehr}]{Bernhard2006Nature}
Bernhard, H., Fischbacher, U., Fehr, E., 2006. {Parochial altruism in humans}.
  Nature 442, 912--915.

\bibitem[{Boyd et~al.(2003)Boyd, Gintis, Bowles, and Richerson}]{Boyd2003PNAS}
Boyd, R., Gintis, H., Bowles, S., Richerson, P.~J., 2003. {The evolution of
  altruistic punishment}. Proc. Natl. Acad. Sci. USA 100, 3531--3535.

\bibitem[{Brandt and Sigmund(2005)}]{Brandt2005PNAS}
Brandt, H., Sigmund, K., 2005. {Indirect reciprocity, image scoring, and moral
  hazard}. Proc. Natl. Acad. Sci. USA 102, 2666--2670.

\bibitem[{Brandt and Sigmund(2006)}]{Brandt2006jtb}
Brandt, H., Sigmund, K., 2006. The good, the bad and the discriminator ---
  errors in direct and indirect reciprocity. J. Theor. Biol. 239~(2), 183--194.

\bibitem[{Brewer(1999)}]{Brewer1999JSI}
Brewer, M.~B., 1999. {The psychology of prejudice: Ingroup love and outgroup
  hate?} J. Soc. Issues 55, 429--444.

\bibitem[{Brown(2000)}]{Brown2000book}
Brown, R., 2000. Group Processes, second edition. Blackwell Publishing, Malden,
  MA.

\bibitem[{Bshary and Grutter(2006)}]{Bshary2006Nature}
Bshary, R., Grutter, A.~S., 2006. {Image scoring and cooperation in a cleaner
  fish mutualism}. Nature 441~(7096), 975--978.

\bibitem[{Chalub et~al.(2006)Chalub, Santos, and Pacheco}]{Chalub2006jtb}
Chalub, F. A. C.~C., Santos, F.~C., Pacheco, J.~M., 2006. The evolution of
  norms. J. Theor. Biol. 241, 233--240.

\bibitem[{Choi and Bowles(2007)}]{ChoiBowles2007Science}
Choi, J.~K., Bowles, S., 2007. {The coevolution of parochial altruism and war}.
  Science 318, 636--640.

\bibitem[{De~Cremer and van Vugt(1999)}]{DeCremer1999EJSP}
De~Cremer, D., van Vugt, M., 1999. {Social identification effects in social
  dilemmas: a transformation of motives}. Eur. J. Soc. Psychol. 29, 871--893.

\bibitem[{Dovidio et~al.(2005)Dovidio, Glick, and Rudman}]{Dovidio2005book}
Dovidio, J.~F., Glick, P., Rudman, L.~A. (Eds.), 2005. On the nature of
  prejudice. Blackwell Publishing, Malden, MA.

\bibitem[{Efferson et~al.(2008)Efferson, Lalive, and
  Fehr}]{Efferson2008Science}
Efferson, C., Lalive, R., Fehr, E., 2008. {The coevolution of cultural groups
  and ingroup favoritism}. Science 321, 1844--1849.

\bibitem[{Fortunato(2010)}]{Fortunato2010PhysRep}
Fortunato, S., 2010. Community detection in graphs. Phys. Rep. 486, 75--174.

\bibitem[{Fowler and Kam(2007)}]{Fowler2007JPoli}
Fowler, J.~H., Kam, C.~D., 2007. {Beyond the self: social identity, altruism,
  and political participation}. J. Politics 69, 813--827.

\bibitem[{Garc{\'\i}a and van~den Bergh(2011)}]{Garcia2011EHB}
Garc{\'\i}a, J., van~den Bergh, J. C. J.~M., 2011. {Evolution of parochial
  altruism by multilevel selection}. Evol. Human Behav. 32, 277--287.

\bibitem[{Goette et~al.(2006)Goette, Huffman, and Meier}]{Goette2006AER}
Goette, L., Huffman, D., Meier, S., 2006. {The impact of group membership on
  cooperation and norm enforcement: Evidence using random assignment to real
  social groups}. Amer. Econ. Rev. 96, 212--216.

\bibitem[{Hewstone et~al.(2002)Hewstone, Rubin, and Willis}]{Hewstone2002ARP}
Hewstone, M., Rubin, M., Willis, H., 2002. {Intergroup bias}. Annu. Rev.
  Psychol. 53, 575--604.

\bibitem[{Ihara(2011)}]{Ihara2011PhilB}
Ihara, Y., 2011. {Evolution of culture-dependent discriminate sociality: a
  gene-culture coevolutionary model}. Phil. Trans. R. Soc. Lond. B 366,
  889--900.

\bibitem[{Jansen and van Baalen(2006)}]{Jansen2006Nature}
Jansen, V. A.~A., van Baalen, M., 2006. {Altruism through beard
  chromodynamics}. Nature 440, 663--666.

\bibitem[{Jones et~al.(1981)Jones, Wood, and Quattrone}]{Jones1981PSPB}
Jones, E.~E., Wood, G.~C., Quattrone, G.~A., 1981. {Perceived variability of
  personal characteristics in in-groups and out-groups: the role of knowledge
  and evaluation}. Person. Soc. Psychol. Bull. 7, 523--528.

\bibitem[{Leimar and Hammerstein(2001)}]{Leimar2001RoyalB}
Leimar, O., Hammerstein, P., 2001. Evolution of cooperation through indirect
  reciprocity. Proc. R. Soc. B 268, 745--753.

\bibitem[{Lize et~al.(2006)Lize, Carval, Cortesero, Fournet, and
  Poinsot}]{Lize2006RoyalB}
Lize, A., Carval, D., Cortesero, A.~M., Fournet, S., Poinsot, D., 2006. {Kin
  discrimination and altruism in the larvae of a solitary insect}. Proc. R.
  Soc. B 273, 2381--2386.

\bibitem[{Masuda and Ohtsuki(2007)}]{MasudaOhtsuki2007RoyalB}
Masuda, N., Ohtsuki, H., 2007. Tag-based indirect reciprocity by incomplete
  social information. Proc. R. Soc. B 274, 689--695.

\bibitem[{Mifune et~al.(2010)Mifune, Hashimoto, and Yamagishi}]{Mifune2010EHB}
Mifune, N., Hashimoto, H., Yamagishi, T., 2010. {Altruism toward in-group
  members as a reputation mechanism}. Evol. Human Behav. 31, 109--117.

\bibitem[{Nakamura and Masuda(2011)}]{NakamuraMasuda2011PLoSComputBiol}
Nakamura, M., Masuda, N., 2011. {Indirect reciprocity under incomplete
  observation}. PLoS Comput. Biol. 7, e1002113.

\bibitem[{Nowak(2006{\natexlab{a}})}]{Nowak2006book}
Nowak, M.~A., 2006{\natexlab{a}}. Evolutionary Dynamics. The Belknap Press of
  Harvard University Press, MA.

\bibitem[{Nowak(2006{\natexlab{b}})}]{Nowak2006Science}
Nowak, M.~A., 2006{\natexlab{b}}. Five rules for the evolution of cooperation.
  Science 314, 1560--1563.

\bibitem[{Nowak and Sigmund(1998{\natexlab{a}})}]{Nowak1998jtb}
Nowak, M.~A., Sigmund, K., 1998{\natexlab{a}}. The dynamics of indirect
  reciprocity. J. Theor. Biol. 194, 561--574.

\bibitem[{Nowak and Sigmund(1998{\natexlab{b}})}]{Nowak1998nature}
Nowak, M.~A., Sigmund, K., 1998{\natexlab{b}}. Evolution of indirect
  reciprocity by image scoring. Nature 393, 573--577.

\bibitem[{Nowak and Sigmund(2005)}]{Nowak2005Nature}
Nowak, M.~A., Sigmund, K., 2005. Evolution of indirect reciprocity. Nature 437,
  1291--1298.

\bibitem[{Ohtsuki and Iwasa(2004)}]{OhtsukiIwasa2004jtb}
Ohtsuki, H., Iwasa, Y., 2004. How should we define goodness?--reputation
  dynamics in indirect reciprocity. J. Theor. Biol. 231, 107--120.

\bibitem[{Ohtsuki and Iwasa(2007)}]{OhtsukiIwasa2007jtb}
Ohtsuki, H., Iwasa, Y., 2007. Global analyses of evolutionary dynamics and
  exhaustive search for social norms that maintain cooperation by reputation.
  J. Theor. Biol. 244, 518--531.

\bibitem[{Ohtsuki et~al.(2009)Ohtsuki, Iwasa, and Nowak}]{Ohtsuki2009Nature}
Ohtsuki, H., Iwasa, Y., Nowak, M.~A., 2009. {Indirect reciprocity provides only
  a narrow margin of efficiency for costly punishment}. Nature 457, 79--82.

\bibitem[{Ostrom and Sedikides(1992)}]{Ostrom1992PsychBull}
Ostrom, T.~M., Sedikides, C., 1992. {Out-group homogeneity effects in natural
  and minimal groups}. Psychol. Bull. 112, 536--552.

\bibitem[{Pacheco et~al.(2006)Pacheco, Santos, and
  Chalub}]{Pacheco2006PlosComputBiol}
Pacheco, J.~M., Santos, F.~C., Chalub, F. A. C.~C., 2006. Stern-judging: A
  simple, successful norm which promotes cooperation under indirect
  reciprocity. PLoS Comput. Biol. 2, 1634--1638.

\bibitem[{Rand et~al.(2009)Rand, Pfeiffer, Dreber, Sheketoff, Wernerfelt, and
  Benkler}]{Rand2009PNAS}
Rand, D.~G., Pfeiffer, T., Dreber, A., Sheketoff, R.~W., Wernerfelt, N.~C.,
  Benkler, Y., 2009. {Dynamic remodeling of in-group bias during the 2008
  presidential election}. Proc. Natl. Acad. Sci. USA 106, 6187--6191.

\bibitem[{Sedikides et~al.(1998)Sedikides, Schopler, and
  Insko}]{Sedikides1998book}
Sedikides, G., Schopler, J., Insko, C.~A. (Eds.), 1998. Intergroup Cognition
  and Intergroup Behavior. Psychology Press, New York.

\bibitem[{Shinada et~al.(2004)Shinada, Yamagishi, and Ohmura}]{Shinada2004EHB}
Shinada, M., Yamagishi, T., Ohmura, Y., 2004. {False friends are worse than
  bitter enemies: ``Altruistic'' punishment of in-group members}. Evol. Human
  Behav. 25, 379--393.

\bibitem[{Sigmund et~al.(2001)Sigmund, Hauert, and Nowak}]{Sigmund2001PNAS}
Sigmund, K., Hauert, C., Nowak, M.~A., 2001. Reward and punishment. Proc. Natl.
  Acad. Sci. USA 98, 10757--10762.

\bibitem[{Suzuki and Toquenaga(2005)}]{SuzukiToquenaga2005JTB}
Suzuki, Y., Toquenaga, Y., 2005. {Effects of information and group structure on
  evolution of altruism: analysis of two-score model by covariance and
  contextual analyses}. J. Theor. Biol. 232, 191--201.

\bibitem[{Traulsen(2008)}]{Traulsen2008EPJB}
Traulsen, A., 2008. {Mechanisms for similarity based cooperation}. Eur. Phys.
  J. B 63, 363--371.

\bibitem[{Traulsen and Nowak(2007)}]{Traulsen2007PlosOne}
Traulsen, A., Nowak, M.~A., 2007. {Chromodynamics of cooperation in finite
  populations}. PLoS ONE 2, e270.

\bibitem[{Trivers(1971)}]{Trivers1971}
Trivers, R.~L., 1971. The evolution of reciprocal altruism. Q. Rev. Biol. 46,
  35--57.

\bibitem[{West et~al.(2007)West, Griffin, and Gardner}]{West2007JEB}
West, S.~A., Griffin, A.~S., Gardner, A., 2007. {Social semantics: altruism,
  cooperation, mutualism, strong reciprocity and group selection}. J. Evol.
  Biol. 20, 415--432.

\bibitem[{West et~al.(2008)West, Griffin, and Gardner}]{West2008JEB}
West, S.~A., Griffin, A.~S., Gardner, A., 2008. {Social semantics: how useful
  has group selection been?} J. Evol. Biol. 21, 374--385.

\bibitem[{Yamagishi et~al.(1999)Yamagishi, Jin, and
  Kiyonari}]{Yamagishi1999chapter}
Yamagishi, T., Jin, N., Kiyonari, T., 1999. {Bounded generalized reciprocity
  --- ingroup boasting and ingroup favoritism}. In: Advances in group processes
  16, 161--197.

\bibitem[{Yamagishi et~al.(1998)Yamagishi, Jin, and Miller}]{Yamagishi1998AJSP}
Yamagishi, T., Jin, N., Miller, A.~S., 1998. {In-group bias and culture of
  collectivism}. Asian J. Soc. Psychol. 1, 315--328.

\bibitem[{Yamagishi and Mifune(2008)}]{YamagishiMifune2008RS}
Yamagishi, T., Mifune, N., 2008. {Does shared group membership promote
  altruism?} Rationality and Society 20, 5--30.

\end{thebibliography}
\end{document}